\documentclass[nohyper,article,11pt,notoc]{JHEP3}
\pdfoutput=1
\usepackage{graphicx}
\usepackage{amssymb}
\usepackage{amsmath}
\usepackage{amsfonts}
\usepackage{latexsym}
\usepackage[all]{xy}
\usepackage{slashed}
\usepackage{slashbox}

\preprint{}

\newcommand{\lsim}{ \mathop{}_{\textstyle \sim}^{\textstyle <}}

\newcommand{\keV}{~\mathrm{keV}}
\newcommand{\keVr}{~\mathrm{keVr}}
\newcommand{\MeV}{~\mathrm{MeV}}
\newcommand{\GeV}{~\mathrm{GeV}}

\newcommand{\cm}{~\mathrm{cm}}
\newcommand{\km}{~\mathrm{km}}
\newcommand{\vesc}{v_{\rm esc}}

\newcommand{\mN}{m_{\scriptscriptstyle N}}
\newcommand{\mX}{m_{\scriptscriptstyle \chi}}
\newcommand{\sX}{\sigma_{\rm n}}

\newcommand{\ER}{E_{\scriptscriptstyle R}}

\newcommand{\be}{\begin{eqnarray}}
\newcommand{\ee}{\end{eqnarray}}

\preprint{MCTP-10-16}
\title{CoGeNT Interpretations}
\author{Spencer Chang$^{(a)}$,   Jia Liu$^{(b)}$, Aaron Pierce$^{(c)}$, Neal Weiner$^{(b)}$,  and Itay
  Yavin$^{(b)}$\\ \it{(a) Physics Department, University of California Davis, Davis, California 95616}  \\ \it{(b) Center for Cosmology and Particle Physics, Department of Physics, New York University, New York, NY 10003}\\ \it{(c) Michigan Center for Theoretical Physics (MCTP) Department of Physics, University of Michigan, Ann Arbor, MI 48109 }} 

\abstract{Recently, the CoGeNT experiment has reported events in excess of expected background. We analyze dark matter scenarios which can potentially explain this signal.  Under the standard case of spin independent scattering with equal couplings to protons and neutrons, we find significant tensions with existing constraints. Consistency with these limits is possible if a large fraction of the putative signal events is coming from an additional source of experimental background. In this case, dark matter recoils cannot be said to explain the excess, but are consistent with it.  We also investigate modifications to dark matter scattering that can evade the null experiments.  In particular, we explore generalized spin independent couplings to protons and neutrons, spin dependent couplings, momentum dependent scattering, and inelastic interactions.  We find that some of these generalizations can explain most of the CoGeNT events without violation of other constraints. Generalized couplings with some momentum dependence, allows further consistency with the DAMA modulation signal, realizing a scenario where both CoGeNT and DAMA signals are coming from dark matter. A model with dark matter interacting and annihilating into a new light boson can realize most of the scenarios considered.
 }

\begin{document}

\section{Introduction}

Numerous experiments have been designed to search for the dark matter (DM), which constitutes the vast majority of all matter.  Most searches have only placed limits.  How do these null results influence the interpretation of a putative signal?  This depends upon  the properties of Weakly Interacting Particles (WIMPs)  themselves.
Variations in the couplings of WIMPs to protons, neutrons and spin \cite{Jungman:1995df}, as well as possible inelastic \cite{iDM,iDMUpdate} or momentum dependent \cite{MDDM,FFDM}  scatterings can all change expectations for signal rates.  They also impact comparisons between different experiments.   In addition, differences in target compositions and energy thresholds cause astrophysical unknowns, such as the local velocity distribution of WIMPs to impact experiments in different ways.  When confronting a new signal, a broad exploration of these ideas is important to discern whether it can be of DM origin, or simply an unexpected background.

Recently, the CoGeNT collaboration reported on a low energy ionization spectrum not immediately identifiable with background~\cite{Aalseth:2010vx}. One possible explanation for these low energy events is a genuine signal from DM nuclear recoils against the germanium target. However, since CoGeNT does not discriminate between nuclear and electron recoils, one must be wary of unexpected backgrounds.   It is thus important to determine the viability of a DM origin of the signal for various scenarios, given other experimental limits.  

Initial work in this direction has already appeared in \cite{Fitzpatrick:2010em} and the updated appendix of \cite{Kopp:2009qt}.  In finding a DM fit,  \cite{Fitzpatrick:2010em} allows the possibility of a substantial contribution to the CoGeNT data from both DM and background, while \cite{Kopp:2009qt} allows only a DM contribution to CoGeNT.  They also differ in their treatment of possible systematic errors associated with the XENON experiment.   Reference \cite{Fitzpatrick:2010em} entertains the possibility that systematic errors at the XENON experiment might degrade their published limits.  In this work, we further explore the viability of DM recoils as an explanation to the CoGeNT data.  We place an emphasis on the effect of the inclusion of background.

In the absence of an exponential background which makes a significant contribution to the data, an interpretation of the CoGeNT data as spin-independent scattering is strongly constrained.  With equal couplings to protons and neutrons,  the CoGeNT $99\%$ confidence region is entirely excluded by recent results from CDMS silicon (CDMS-Si) run~\cite{Filippini}. We demonstrate that parameter space can open up for other relations between these couplings.  We also examine other scenarios, such as light inelastic DM (iDM), spin-dependent couplings, and momentum dependent interactions. Aside from elastic spin-dependent interactions, we find that these generalizations are viable interpretations for CoGeNT not excluded by other experiments.

These analyses can be sensitive to the inclusion of backgrounds. Therefore we consider the effect of an additional exponential background component at low energies. Depending on the amount of background allowed, one lessens the tension with CDMS-Si.   However, completely avoiding the Si constraints requires a large amount of background, with at least one bin below the known background sources containing more than 50\% of its events from a new background source. In this case,  the CoGeNT data is not really ``explained" by DM recoil events, rather it would be fair to say the data is not {\em inconsistent} with a DM signal.  

 We also investigate whether the preferred regions for the DAMA modulation signal can be made consistent with CoGeNT.  Typically, the DAMA region requires a significantly lower cross section than CoGeNT, however this assumes the level of channeling that the DAMA collaboration has claimed from Monte Carlo simulations \cite{Bernabei:2007hw}.  In some models, it is possible to get consistency between them.  This typically requires lowering the CoGeNT DM contribution by introducing additional background and/or reducing the amount of channeling in NaI crystals to raise the required cross section for DAMA.  For certain specialized cases, it is even possible to get consistency by only moving DAMA up by a reduction in channeling.  This raises the possibility of consistency with all constraints, while simultaneously explaining CoGeNT and DAMA as true DM signals.

The outline for the rest of the paper is as follows.  In Section \ref{sec:methodology}, we discuss how we interpret the potential signals of CoGeNT and DAMA and how we derive constraints from the other direct detection experiments that impact light mass WIMPs.  Those who are interested in our main results can skip to Section \ref{sec:SI}, where we begin our discussion on whether WIMPs which scatter with spin independent interactions can explain the data.  We will find that there is tension between the signals and constraints for the standard assumption of equal proton and neutron couplings. The inclusion of a background opens regions of parameter space, but the points in these regions generally fail to explain at least half of the events in some bin below 1 keVee.  In Section \ref{sec:alternatives}, we discuss alternative scenarios, starting with generalized couplings for protons and neutrons, spin dependent scattering, moving on to momentum dependent interactions and finally, inelastic DM.  For these scenarios, we find that generalized couplings along with  momentum dependent interactions are the most promising in terms of explaining all experiments.  We also comment on the types of models which might explain the necessary couplings and interactions. Models where the DM annihilates into and interacts via a new light boson provide simple explanations for most scenarios we consider.  Finally, in Section \ref{sec:conclusions}, we conclude.  
 
\section{Methodology \label{sec:methodology}}
For DM scattering, the recoil rate at direct detection experiments can be written as
\begin{eqnarray}
\frac{dR}{dE_R}= \frac{N_T \mN \rho_\chi}{2\mX \mu^2} \sigma(q^2) \int^\infty_{v_{min}} \frac{f(v)}{v} dv.
\end{eqnarray}
Here $N_T$ is the number of target nuclei per unit mass of the detector, $\mN$ is the nucleus mass, $\mu$ is the reduced mass of the WIMP/target nucleus system, and $\rho_\chi$ is the local DM mass density\footnote{For the rest of this paper we will choose $\rho_\chi = 0.3\GeV/\cm^3$, but see~\cite{Salucci:2010qr} for a recent discussion of this quantity. The precise value does not modify any of the comparisons between experiments made below since it is a common factor affecting all the different experiments in the same manner.}, $\mX$ is the DM mass, $\sigma(q^2)$ is the DM-nucleus momentum dependent cross-section, $q = \sqrt{2\mN \ER}$, and $f(v)$ is the local DM speed distribution. 

For spin independent (SI) interactions, we have 
\begin{equation}
\sigma_{\rm SI}(q^2) = \frac{4 G_F^2 \mu^2}{\pi}  \left[Z f_p+(A-Z)f_n\right]^2 F^2(q^2),
\end{equation}
where $G_F$ is the Fermi constant, $f_p,f_n$ are respectively the couplings to the proton and neutron, and $F^2(q^2)$ is the form factor.  Specifically, we use the Helm form factor \cite{Lewin:1995rx}.  The cross section on nucleons is $\sigma_{(p,n)} = \frac{4}{\pi} G_F^2 \mu_{(p,n)}^2 f_{(p,n)}^2$, where $\mu_{(p,n)}$ is the reduced mass of the WIMP-nucleon system.
For spin dependent (SD) interactions, we have
\begin{eqnarray}
\sigma_{\rm SD}(q^2)& =& \frac{32 G_F^2 \mu^2}{2J+1} [a_p^2 S_{pp}(q^2)+a_p a_n S_{pn}(q^2)
+ a_n^2 S_{nn}(q^2) ],
\end{eqnarray}
where $a_p,a_n$ are respectively the couplings to the proton and neutron, and the $S$ factors are the form factors for SD scattering.  The corresponding nucleon cross sections are $\sigma_{(p,n)} = \frac{24}{\pi}G_F^2 \mu_{(p,n)}^2 a_{(p,n)}^2$.

\subsection{CoGeNT}
CoGeNT is a germanium detector with extremely low noise and excellent energy resolution, making it particularly sensitive to light WIMP scattering.  In their latest release  \cite{Aalseth:2010vx}, they see an interesting feature at low energy, consistent with the exponential spectra expected from DM scattering.
The low energy spectrum ($0.37 < E ({\rm keVee}) < 3.2 $) reported by CoGeNT consists of 57 energy bins. After unfolding the efficiency,  the very first energy bin, $0.348 < E ({\rm keVee}) < 0.398$ has close to 100 events, considerably more than any of the other bins. None of the models described below result in a reasonable fit to this point. It is particularly sensitive to the very low efficiency at this energy as well as any possible background associated with the energy threshold. We therefore discard it from the rest of the analysis. We pair the remaining bins and are left with a total of 28 bins which are used throughout the analysis below.  The observed counts in this binning can be seen in Fig.~\ref{fig:nobgfit}. 

 We now move on to describe the models used to fit the data, which were also used in \cite{Aalseth:2010vx}. In the simplest case, we assume that the background is made up entirely of a constant component as well as a component describing the L-shell energy levels associated with electron capture in $^{68}{\rm Ge}$ and $~^{65}{\rm Zn}$,
\begin{equation}
\label{eqn:simplebg}
{\rm BG}_1(E; c_0,c_1) = c_0 + c_1 \left( \exp\left(-\frac{(E-1.298)^2}{0.0975^2} \right) + 0.4 \exp\left( -\frac{(E-1.1)^2}{0.0975^2}\right) \right),
\end{equation}
where $c_0$ and $c_1$ are unknown positive coefficients that are allowed to float in the fit. The 0.4 indicates the expected relative strength of these two lines, and the $0.0975~{\rm keVee}$ is the energy resolution~\cite{collar}. The width is set by the detector resolution. It is important to realize that the constant background does well in explaining the data between $1.5 \lesssim E ({\rm keVee}) < 3.2 $, whereas the peaks are described by the double Gaussian. It is therefore only the first $\sim 5$ bins which deserve explanation\footnote{In some of the examples discussed below, the fit through the 6$^{th}$ and 8$^{th}$ is somewhat lacking. However, we refrain from making any inferences about the WIMP contribution based on these bins since they are likely associated with the L-shell peaks.}. As we shall see in the next section, these bins can be explained by a model of WIMP nuclear recoils which consist of two parameters: the WIMP mass, $\mX$, and the WIMP-nucleon cross-section, $\sX$.

We also consider a more complicated background model including an additional component (of as yet an unknown origin) with an exponential shape,
\begin{equation}
\label{eqn:expbg}
{\rm BG}_2(E; c_0,c_1,c_2,c_3) = {\rm BG}_1(E; c_0,c_1) + c_2 \exp(-c_3 E),
\end{equation}
where $c_i$ are again all positive coefficients to be determined from the fit.  The CoGeNT analysis \cite{Aalseth:2010vx} suggests that such backgrounds can come from sources such as surface events or unvetoed neutrons, although they find it hard to accommodate the size required.  As we shall see in detail below, this additional contribution allows for more freedom in accounting for the low energy events and consequently opens up a larger region in the WIMP parameter space, $\mX$-$\sX$. As emphasized above, the constant background together with the L-shell Gaussians account well for the data throughout except for the first 5 bins. Therefore, it is important to recognize that in this more complicated model the two parameters describing the exponential background together with the two parameters describing the WIMP nuclear recoil are effectively responsible for only 5 data points. As we shall see below, this is not the only difficulty with this model. 

To calculate the WIMP signal we assume a Lindhart type quenching factor for CoGeNT $\ER({\rm keVee}) = Q\times \ER({\rm keV})^{1.1204}$, $Q=0.19935$. Therefore, the events below $\lesssim 1~{\rm keVee}$ correspond to nuclear recoil energies $\lesssim 5\keV$. For a larger $Q$, the same $E_{\rm ee}$ corresponds to a lower $E_{\rm R}$, allowing lower mass particle to contribute to the signal. Indeed, with a larger Q the CoGeNT region in Fig. \ref{fig:nobgcontours} moves to lower DM mass. With a smaller Q, it moves to higher DM mass. But the change is mild with Q in the range $[0.157, 0.22]$, a range apparently consistent with \cite{Lin:2007ka, Barbeau:2007qi}. 
As we will see below, the null results from other experiments, particularly CDMS Si, would prefer a lower mass.  A more extreme value of $Q=0.25$ might make CoGeNT consistent (90\% CL) with the CDMS Si bound by moving part of the allowed region to $m=6.5$ GeV. 
 
When forming confidence intervals in the WIMP parameter space below, we use the likelihood profile method to eliminate the nuisance parameters associated with the background~\cite{Cowan:2008zza}. This is done by first minimizing the $\chi^2$ function (defined over the 28 bins) with respect to the background parameters, $c_i$, at every point in the $\mX - \sX$ space while keeping $\mX$ and $\sX$ fixed,
\begin{equation}
\label{eqn:profile}
\chi^2(\mX, \sX) = {\rm min}|_{c_i}~\chi^2(\mX,\sX,c_i).
\end{equation}
The global minimum is then found by minimizing $\chi^2(\mX, \sX)$. Finally, the confidence interval is constructed by solving
\begin{equation}
\chi^2(\mX, \sX) - \chi_{\rm min}^2 = \Delta \chi^2
\end{equation}
for the appropriate $\Delta\chi^2$ ($\Delta\chi^2 = 4.61 (9.21)$ for the 90\% (99\%) confidence level for two variables). Since it is part of our goal to understand the relative importance of the exponential background component of Eq.~(\ref{eqn:expbg}), we aim to restrict its contribution to the fit in the low energy bins. We achieve that by performing the minimization in Eq.~(\ref{eqn:profile}) with the constraints that the exponential background is not responsible for more than a certain fraction of the events in each of the first 5 bins. 

\subsection{DAMA}
Over the past $\sim 10$ years, DAMA/NaI and its upgrade DAMA/LIBRA has observed an annual modulation effect which is potentially due to DM.  Light DM near the 10 GeV range has been proposed to explain the DAMA signal while remaining consistent with other experiments \cite{Gelmini:2004gm}.  Recent analyses show that channeling \cite{Bernabei:2007hw} is required to evade other constraints \cite{Petriello:2008jj,Fairbairn:2008gz,Bottino:2008mf,Savage:2008er,Feng:2008dz}, though the details of the observed DAMA spectrum (in particular the decrease in modulation seen in the lowest bin) constrain even this interpretation  \cite{Chang:2008xa}. 
In this analysis, we use the data from their latest release which includes the last two years of DAMA/LIBRA data~\cite{Bernabei:2010mq}, which has improved error bars from the statistics.  

We follow the analysis as described in \cite{Chang:2008xa}.  We take into account the channeling effect as parameterized in~\cite{Foot} which is based on the predictions from the DAMA collaboration~\cite{Bernabei:2007hw}.  When we later discuss reduced channeling, we mean an overall factor multiplying these predicted channeling probabilities.   We perform a $\chi^2$ fit to their extracted modulation spectra with signal alone, using the first 8 bins from 2-6 keVee and add the 6-14 keVee region as a single bin.  Similar to the CoGeNT analysis, we subtract the lowest $\chi^2$ to determine the 90\% and 99\% CL region for the signal parameters.  As was noted in~\cite{Chang:2008xa}, this fit to the spectra restricts the WIMP mass from being too small, generically $> 10$ GeV.  If we treat the first bin with some additional uncertainty or throw it out completely, the lowest mass will decrease by about 1-2 GeV.  However, we caution the reader to keep in mind that our plots {\em do not} take this possibility into account.  Finally, to give an idea for how the DAMA preferred region changes if the channeling is smaller than the prediction in~\cite{Bernabei:2007hw}, we will also consider in our plots the $\chi^2$ region when we turn off channeling.  For reduced channeling, the DAMA region will then lie somewhere between these two different $\chi^2$ regions.

\subsection{Constraints}
To determine constraints on the CoGeNT and DAMA fits, we focus on direct detection experiments which are sensitive to WIMPs in the mass range 5-15 GeV.  
The most constraining are the results from the CDMS silicon detectors.  We use the published 2-tower data \cite{Akerib:2005kh} and the 5-tower data shown in \cite{Filippini}, both of which did not see any signal events.  Both runs had thresholds turning on at 7 keV, which we take into account using the efficiencies in \cite{Filippini}.  As a test of the sensitivity to this turn-on, we also considered the effect of using a higher threshold.  An 8 keV threshold gives very little change, while 9 keV generally allow masses .5 GeV larger.  The 9 keV threshold seems too conservative though, since at 8 keV, the acceptance is already 8\%.   Thus, threshold effects seem unlikely to help open the parameter space.  

For XENON10, we use the reanalysis in \cite{Angle:2009xb}, which saw no events down to the stated S2 threshold.  In addition, we apply a conservative limit, by using the newest measurements of $\mathcal{L}_{eff}$ reported in~\cite{Manzur:2009hp}, which increases the energy threshold above the original 4.5 keV.  Aside from the central values of $\mathcal{L}_{eff}$, we also explore varying this within its 1$\sigma$ error bars at low energy, which can increase the energy threshold, significantly weakening the limit for light masses.  

We also take into account the recent results from SIMPLE \cite{Felizardo:2010mi}, which looks for WIMP-induced bubble nucleations in superheated liquid ${\rm C_2 Cl F_5}$.  As their limit, we take the 90\% CL Poisson limit from their four unaccounted events, allowing $\sim$ 8 events.  Not only is SIMPLE sensitive to light mass WIMPs due to its carbon and fluorine, but it has some of the best limits on spin dependent interactions on protons through its fluorine component.  

Finally, we do not display CDMS germanium results~\cite{Ahmed:2009zw}, as they are typically weaker than these other constraints, due to their higher energy threshold.    

\subsection{Dependence on the Dark Matter Velocity Distribution}

The fit to the observed events as well as the exclusion limits derived from null results depend on the velocity distribution chosen for the DM. A qualitative understanding of this dependence is important to account for the inherent uncertainties involved with the velocity distribution as well as to facilitate a comparison between the different experiments and limits.

For elastic scattering, the minimum velocity needed to scatter with a deposited energy $\ER$ is, 
\begin{eqnarray}
\label{eqn:betaminelastic}
v_{\rm min}^2 &=& \frac{\mN \ER}{2\mu^2},
\end{eqnarray}
where $\mN$ is the mass of the target nucleus and $\mu$ is the reduced mass of the WIMP/target nucleus system. Throughout this work, we will assume that DM has a Maxwell-Boltzman velocity distribution with a most probable velocity, $v_0$, and an escape velocity $v_{\rm esc}$.  As representative values, we take $v_0 = 230,270$ km/s and $v_{esc}=500,600$ km/s. We assume that the velocity dispersion is isotropic, although, in general it need not be. In halos with flat rotation curves, the radial velocity dispersion is expected to follow $v_0 = v_{rot}$ \cite{Drukier:1986tm}, thus $v_0 = 230, 270$ km/s are good representative values, with lower values not yielding a better agreement between experiments. The escape velocity can be inferred from studying the velocities of bound objects, and the RAVE survey has indicated that the escape velocity of the galaxy lies within the range  $498 < v_{esc} < 608$ km/s at 90\% confidence \cite{Smith:2006ym}.

The nuclear recoil rate depends on the recoil energy roughly as $\exp(-\ER/E_{max})$, with $E_{max} = 2\mu^2v_0^2/\mN$. For CoGeNT, which uses germanium as a target, $\mu \sim \mX$ for $\mX \sim 10\GeV$. The best fit to the data essentially determines $E_{max}$ and we can expect fits with higher values of $v_0$ to favor lower WIMP masses. The variation in $E_{max}$ with respect to the WIMP mass is also larger for larger $v_0$ and so the confidence interval on the WIMP mass is consequently smaller. Both of these qualitative expectations are borne out in the detailed fits below. The dependence on $v_{\rm esc}$ is similar in nature, but considerably weaker in its effect. 

The limits set by other experiments with null results are also sensitive to the velocity parameters. As mentioned earlier, the most constraining result comes from CDMS-Si. As discussed in the previous paragraph, higher $v_0$ values result in lower implied WIMP mass from the best-fit to the CoGeNT data, and hence one may try to avoid the CDMS-Si constraint by considering DM velocity distributions with higher $v_0$. However, the reduced mass of the WIMP-nucleus system is not very different between germanium and silicon and so higher $v_0$ also implies a corresponding tighter limit on the WIMP mass from CDMS-Si. Similar remarks hold for some of the other exclusion limits. 

Moreover, it is intriguing to note that one can only use velocity distributions to push the limits so far. A simple analysis of kinematics can show this.  From Eq.~(\ref{eqn:betaminelastic}) we note that $v_{\rm min} \approx 403~{\rm km/s}~(10\GeV/\mX)$ for silicon with a threshold at $\ER=7\keV$ as compared with $v_{\rm min} \approx 460~{\rm km/s}~(10\GeV/\mX)$ for germanium at $\ER = 5\keVr$. Therefore, simply assuming that signal is present in the $\sim 1$ keVee $\approx  5$ keVr bin implies that CDMS-Si should be sensitive {\em to the same particles in the halo}, irrespective of the Maxwell-Boltzmann velocity profile parameters.

Thus, while changes in the velocity parameters certainly result in different regions of interest for the WIMP mass, they do not markedly affect the excluded regions in the $\mX-\sX$ parameter space relative to the regions implied by the fit to the CoGeNT data. 

\section{Spin Independent Elastic Scattering \label{sec:SI}}
\subsection{Interpreting the CoGeNT signal assuming no exponential background}

\begin{figure}[h]
\begin{center}
\includegraphics[width=0.45\textwidth]{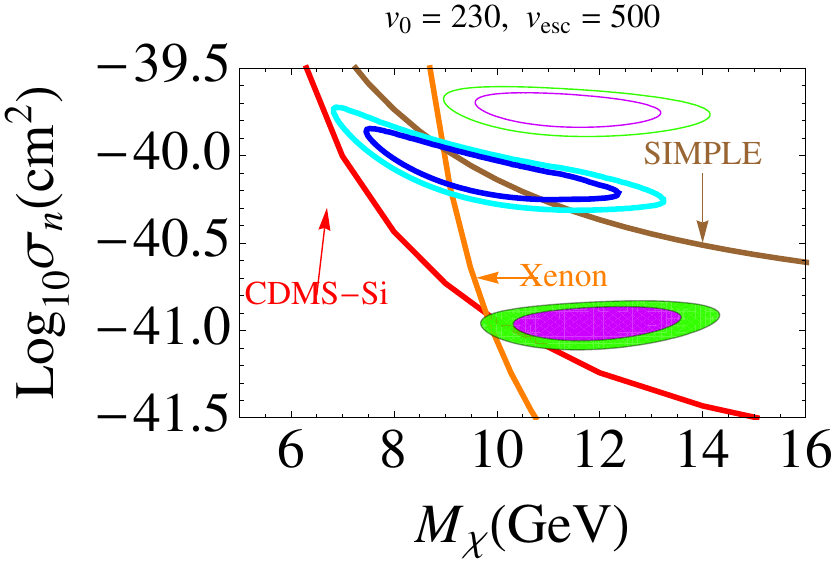}
\includegraphics[width=0.45\textwidth]{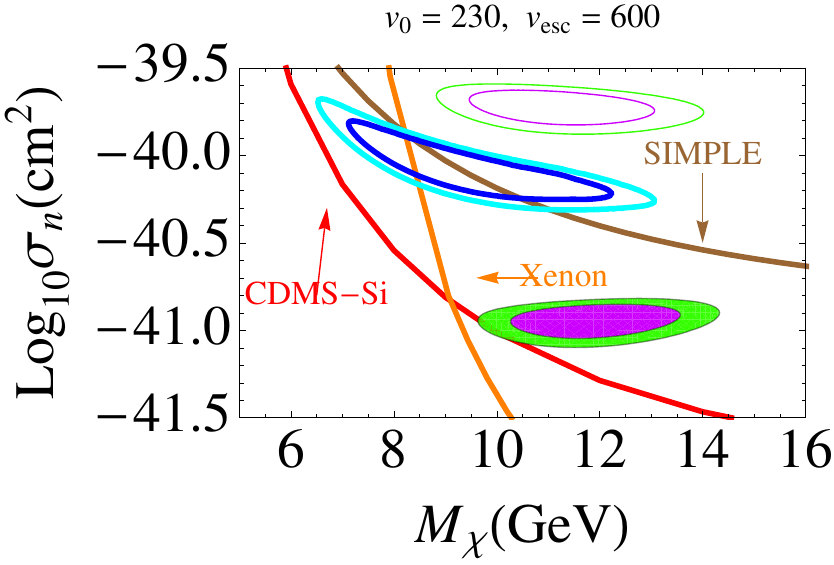}\\
\includegraphics[width=0.45\textwidth]{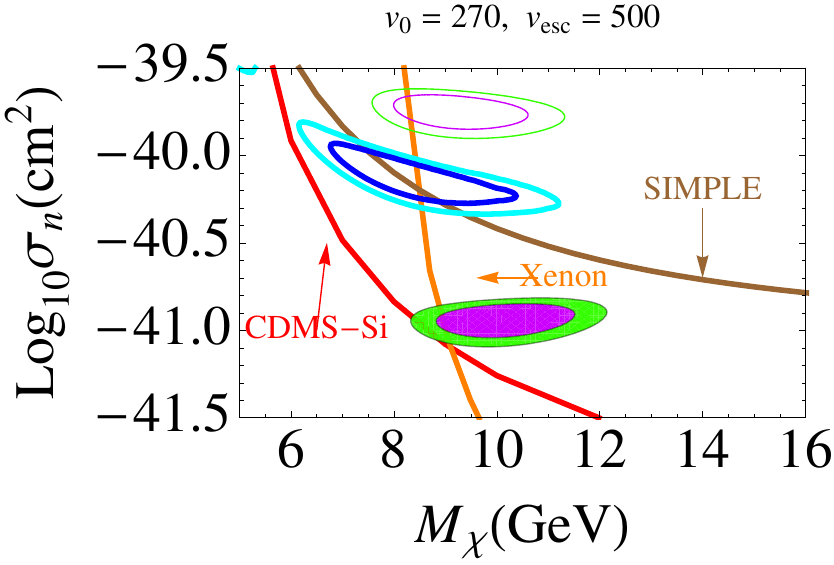}
\includegraphics[width=0.45\textwidth]{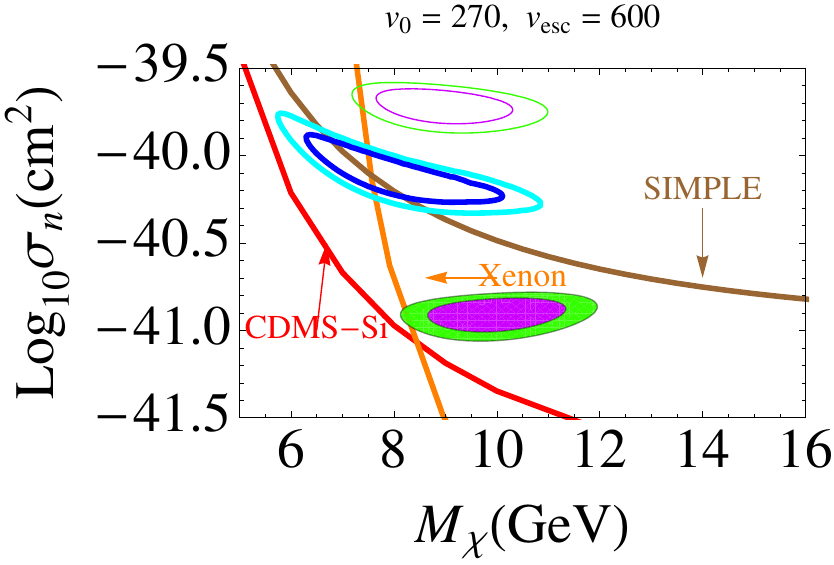}
\end{center}
\caption{SI scattering with equal proton and neutron couplings, $f_p=f_n$.  Plotted are the confidence interval for CoGeNT data shown in blue (90\%) and cyan (99\%), taking the constant and L-shell background Eq.~(\ref{eqn:simplebg}). Also plotted are the 90\% exclusion limits coming from CDMS-Si (red) and XENON10 (orange) and SIMPLE (brown).  
Also plotted are the DAMA confidence intervals shown in magenta (90\%) and green (99\%), where the filled (unfilled) contours assume 100\% (no) channeling.}
\label{fig:nobgcontours}
\end{figure}

The simplest question to consider is whether the low energy ``excess" at CoGeNT can be well explained by elastic WIMP--nucleon scattering, without the assumption of any significant unknown background.  We thus work under the assumption that all the events at low energy are signal events, using the standard case of equal proton and neutron couplings. In this case, we use only the L-shell background form together with a uniform constant background as in Eq. (\ref{eqn:simplebg}). In Fig. \ref{fig:nobgcontours} the confidence intervals for several choices of the velocity distribution parameters are shown. The fits are generally good with $\chi^2 \approx 22$ for $d.o.f = 28 - 4 = 24$ using the four parameters $(\mX,\sX,c_0,c_1)$ with the constraint of positivity imposed on each. Fig. \ref{fig:nobgcontours} also contains the DAMA preferred regions for the case of channeling turned on and off.  
It illustrates the $90\%$ exclusion limits from CDMS-Si, XENON10, and SIMPLE.  In this case, the regions implied by the CoGeNT and DAMA data are ruled out almost entirely by these null results.

\begin{figure}[h]
\begin{center}
\includegraphics{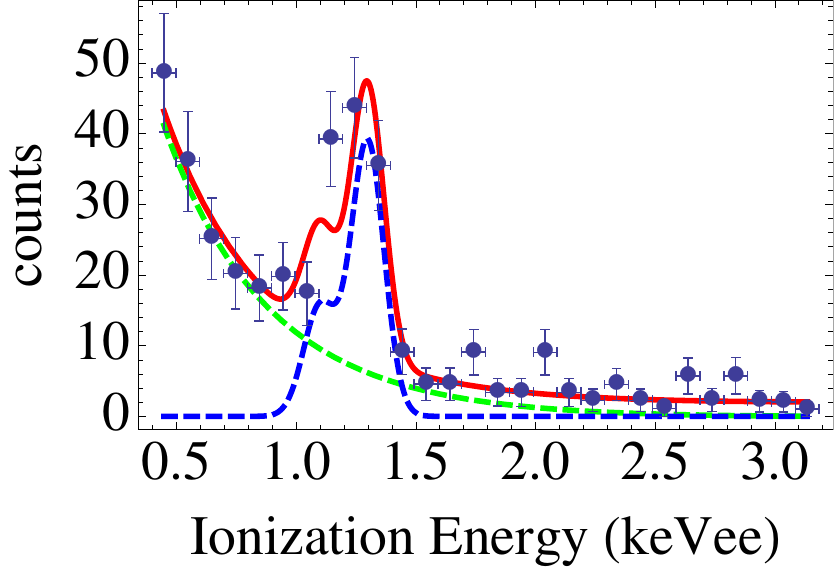}
\end{center}
\caption{Best fit (red-solid) for the low energy CoGeNT data with a WIMP signal component (green light-dashed). The WIMP best-fit parameters are $\mX = 9.4\GeV$ and $\sX = 0.84\times 10^{-40}\cm^2$ and the velocity distribution was taken to be Maxwellian with $v_0 = 230\km/\sec $, and $\vesc = 500\km/\sec$. Notice that the Gaussian peaks from the L-shell background component (blue-dashed) do not contribute to the first 5 energy bins from the left. This part of the spectrum is entirely explained by the WIMP contribution.}
\label{fig:nobgfit}
\end{figure}

In Fig.~\ref{fig:nobgfit} we show the best-fit spectrum for the CoGeNT data using the background of Eq.~(\ref{eqn:simplebg}), supplemented by a WIMP signal with $v_0 = 230\km/\sec $, and $\vesc = 500\km/\sec$.  The WIMP component provides an excellent fit to the low energy excess. The best fit point is at 9.4 GeV, which is well excluded by the CDMS Si data. The exact placement of the CDMS-Si bound is dependent on their acceptance as a function of energy. As discussed earlier, an unreasonably large change would be required to allow consistency between CoGeNT and CDMS-Si. Even if were to obtain consistency between CoGeNT and CDMS-Si, it is hard to see how one might additionally get this region to overlap with DAMA. Increasing the DAMA cross section simply requires the channeling fraction be smaller than supposed.  On the other hand, a shift to lower masses requires modifications to the lowest bins of DAMA, as discussed earlier.

\subsection{Interpreting the CoGeNT signal with an exponential background}
 The analysis in the previous section neglected the very reasonable possibility that an unknown background could appear at low energies, thus opening up  ranges of parameter space not constrained by other experiments. Such an analysis is subtle, however, because it is precisely the fact that the backgrounds are thought to be small that is motivating the DM explanation. One must be careful, therefore, that one does not replace signal with background and then claim those same events as evidence for a DM signal.

\begin{figure}[h]
 \begin{center}
\includegraphics[width=0.45\textwidth]{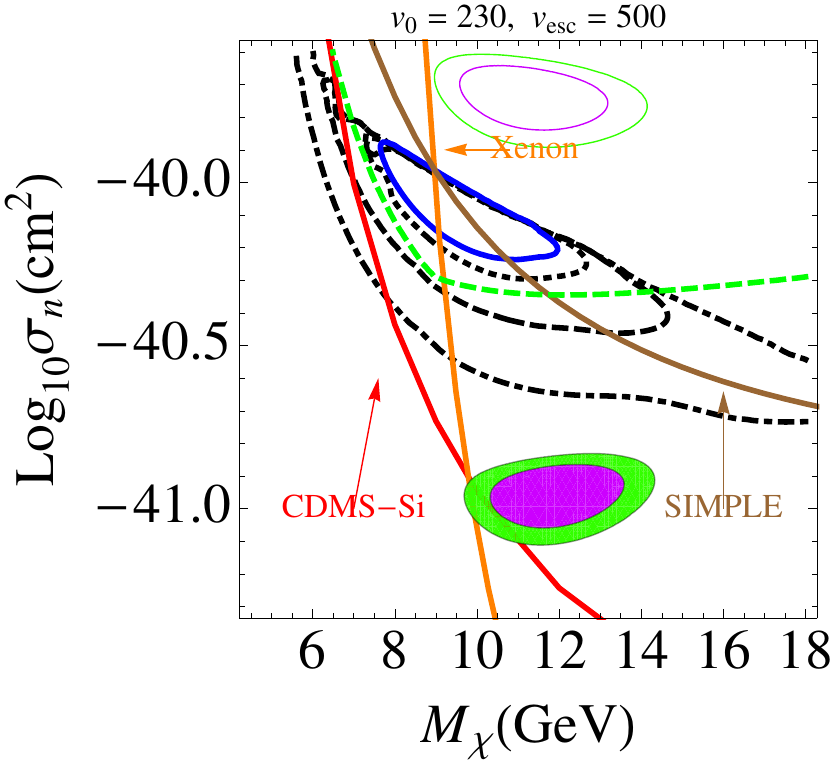}
\includegraphics[width=0.45\textwidth]{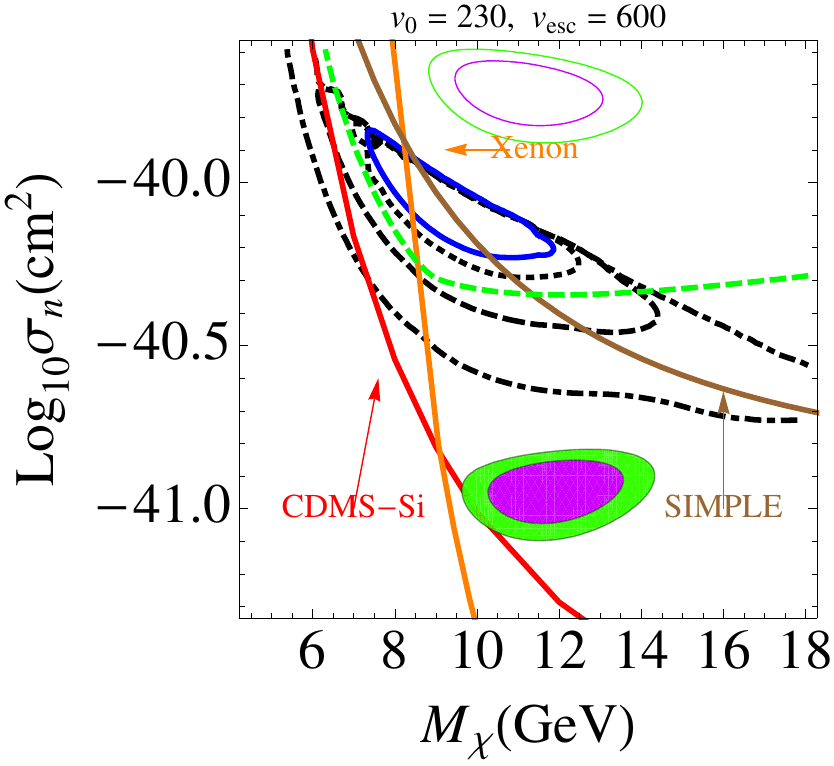}\\
\includegraphics[width=0.45\textwidth]{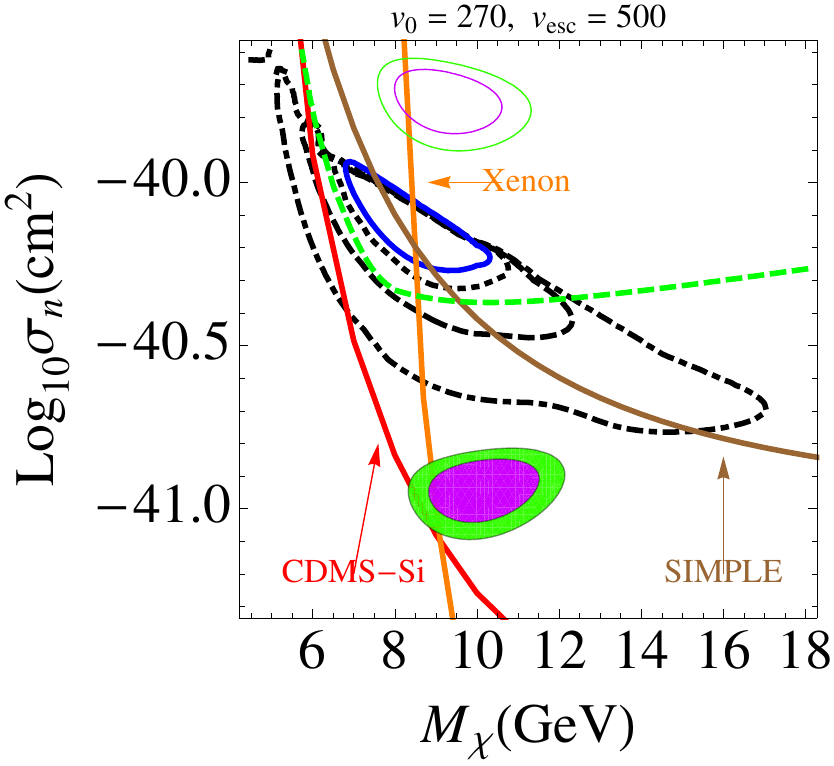}
\includegraphics[width=0.45\textwidth]{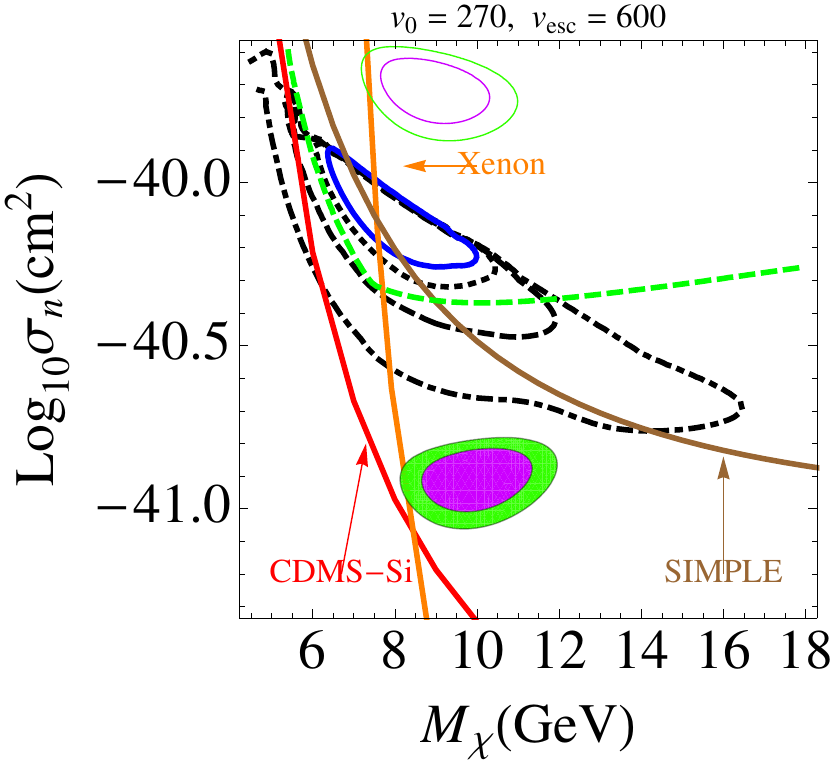}
\end{center}
\caption{Confidence interval for CoGeNT data for $0\%$, $10\%$, $30\%$, and $50\%$ exponential background contribution, solid, dashed, dotted, dot-dashed, respectively. Below the green light-dashed line DM fails to explain 50\% of the signal in at least one bin ({\em see text}).  Other curves and contours are labeled as in figure \ref{fig:nobgcontours}. }
\label{fig:expbgcontours}
\end{figure}

We extend the analysis to allow for some amount of background with an exponential form as in Eq.~(\ref{eqn:expbg}).  Such considerations were previously explored in \cite{Fitzpatrick:2010em}, here we quantify in a different way what is required of the background.  We do not allow this contribution complete freedom, but instead perform a nonlinear $\chi^2$ minimization with the constraint that the exponential part of the background does not exceed $p=10\%,30\%,50\%$ of the number of counts in each of the first 5 bins. The resulting contours are shown in Fig.~\ref{fig:expbgcontours}. These fits are generally good in the sense that $\chi^2_{\rm best~fit}/d.o.f . \sim 1$. As noted earlier, this is mostly due to the fact that most of the data $E \gtrsim 1~{\rm keVee}$ is well fit with Eq.~(\ref{eqn:simplebg}) alone. As shown in the plots,  allowing $p = 50\%$ opens up a part of parameter space not excluded by any other experiment.
 
\begin{figure}[h]
\begin{center}
\includegraphics[width=0.45\textwidth]{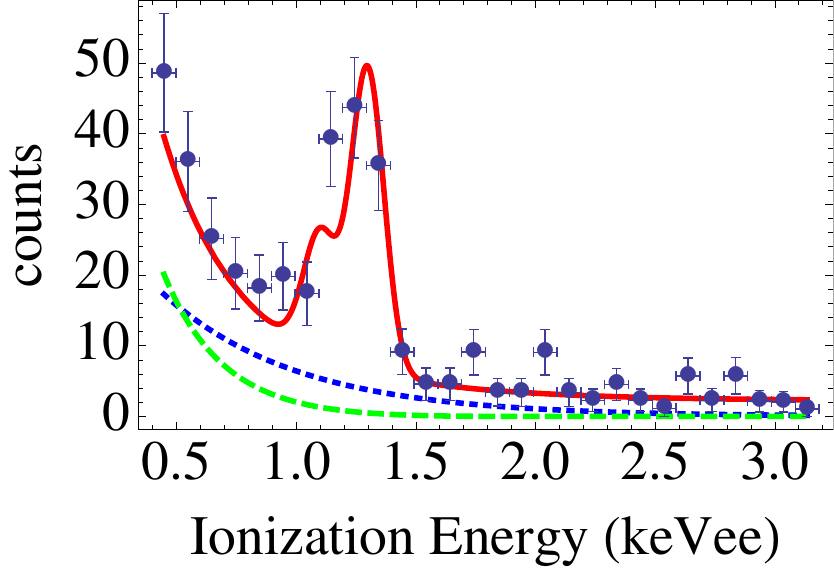}\quad
\includegraphics[width=0.45\textwidth]{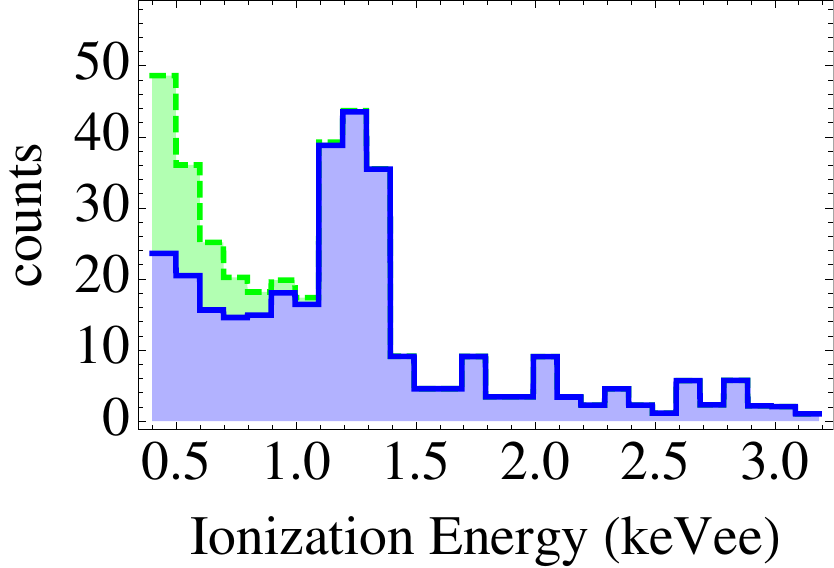}
\end{center}
\caption{On the left we plot a fit (red-solid) for the low energy CoGeNT data with a WIMP signal component (green-dashed) and a  $50\%$ exponential background (blue-dotted). The WIMP parameters are $\mX = 7\GeV$ and $\sX = 0.64\times 10^{-40}\cm^2$ and the velocity distribution was taken to be Maxwellian with $v_0 = 230\km/\sec $, and $\vesc = 500\km/\sec$. On the right, the fit's WIMP component (dashed-green) in each bin is stacked on top of the residual in that bin (solid-blue). The residual is simply the WIMP component subtracted from the data at every bin. It is evident that a very large background (i.e. non-WIMP) component is needed at low energy in order to accommodate such a light WIMP. }
\label{fig:expbgfit}
\end{figure}

 However, it is important to understand how this allowed part of parameter space fits the data. On the left pane of Fig.~\ref{fig:expbgfit} we plot the best-fit to the data for $\mX = 7\GeV$ and $\sX = 0.64\times 10^{-40}\cm^2$ for $v_0 = 230\km/\sec $, and $\vesc = 500\km/\sec$ (not excluded by any other experiment), with the exponential background restricted to be less than $50\%$ of any of the first 5 bins. Notice that the WIMP signal accounts for a substantial part of the first two bins, but almost nothing in the two bins directly to the left of the Gaussian peak. It is the exponential background which fills this gap as much as it can under the restriction of being less than $50\%$ of the content of each bin. Indeed, these bins are underpopulated. So, while this choice of $\mX$ and $\sX$ results in a reasonable fit to the data, it can hardly be said to explain the data. This is illustrated in the right pane of Fig.~\ref{fig:expbgfit} where we plot the WIMP contribution to every energy bin. As is clear from the plot, a substantial amount of background is needed below $E \gtrsim 1~{\rm keVee}$ where the low energy bins directly to the left of the Gaussian peak must be entirely due to background. So, while the constrained background is (by definition) insufficient to produce a good fit for the content of the first 5-6 bins, the WIMP contribution is only improving the fit in the first 3 bins. The few bins directly to the left of the Gaussian peak are left unaccounted for. 
 
We emphasize this point: {\em requiring that the background component is less than 50\% of the events in each bin in a statistical fit does not necessarily imply that even a substantial fraction of every bin is populated by dark-matter events.} Instead, we make the harder requirement - not a statistical one - that DM actually contributes at least 50\% of the observed bin count for each of the first five bins. This motivates the green-dashed lines  in Fig. 3 and following Figs. -- below these lines the WIMP signal fails to explain more than 50\% of the data in any of the first 5 bins. The large regions allowed with additional background components are somewhat misleading - much of the regions leave the events just below the L-shell peaks unexplained.

\section{Alternative Scenarios \label{sec:alternatives}}
It seems challenging to explain the CoGeNT data with a simple WIMP with standard SI elastic scattering, even in the presence of a reasonable background. However, there are many variations on the interactions of WIMPs that are possible: the relative neutron and proton couplings,  couplings to spin, momentum dependent interactions, and inelastic scattering are all simple possibilities that can dramatically alter the relative constraints.

\subsection{Spin Independent Elastic Scattering with Generalized Couplings}

We first consider deviations from the relation $f_{p} = f_{n}$. When compared with the Si at CDMS, the signal(s) at CoGeNT (and DAMA) occur in relatively neutron rich nuclei.  Therefore, coupling choices emphasizing this fact can potentially open a window.  In particular, couplings that nearly obey the relation $f_{p} = -f_{n}$ substantially weaken the bounds from silicon -- the limits from ${}^{28}$Si disappear, leaving only the effects of the subdominant isotopes of ${}^{29}$Si and ${}^{30}$Si, which together compose less than 8$\%$ of the natural abundance.  Near the region $f_{p} = -f_{n}$, the constraints from XENON10, which also has an imbalance in the number of protons and neutrons, remain relevant. 

\begin{figure}[h]
\begin{center}
\includegraphics[width=0.45\textwidth]{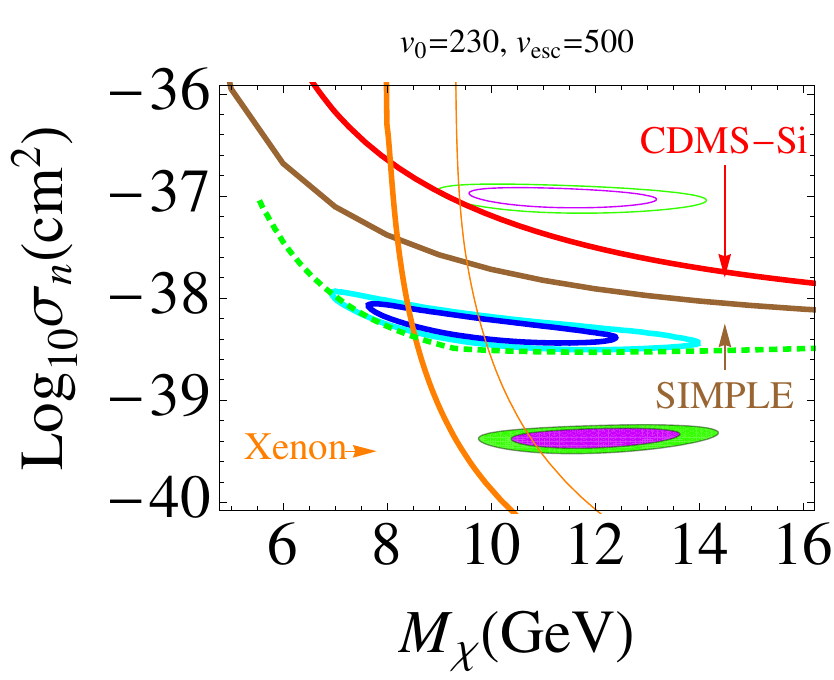}
\includegraphics[width=0.45\textwidth]{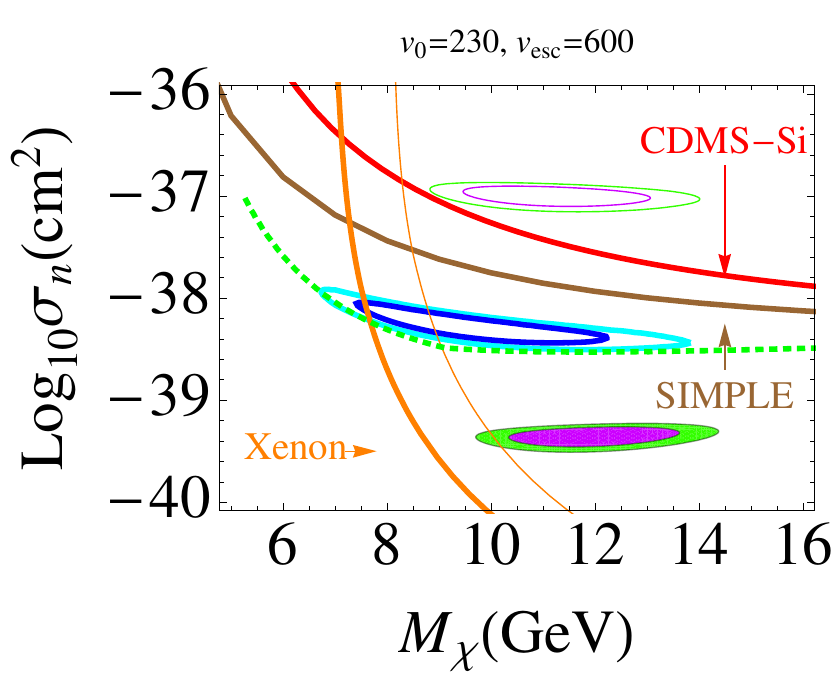}\\
\includegraphics[width=0.45\textwidth]{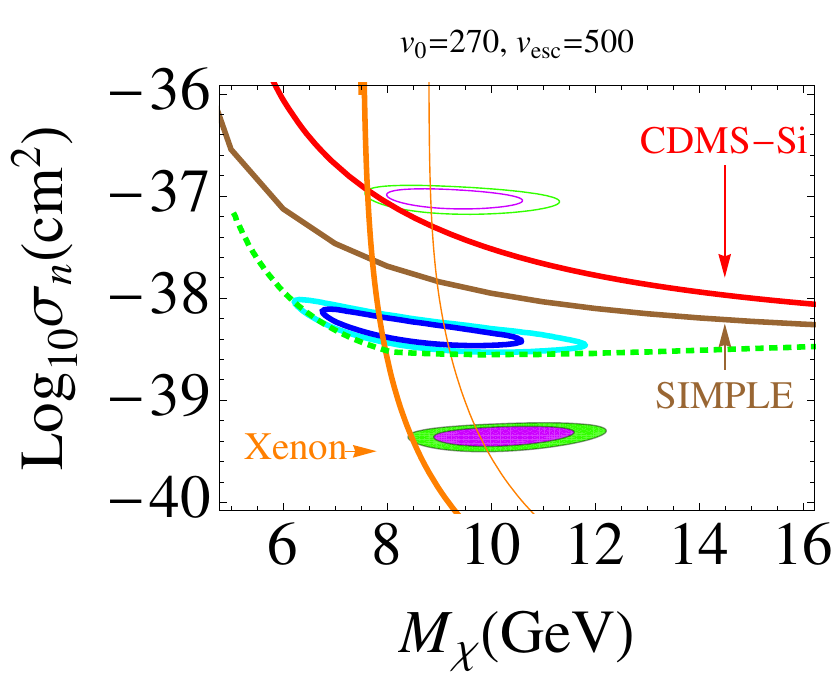}
\includegraphics[width=0.45\textwidth]{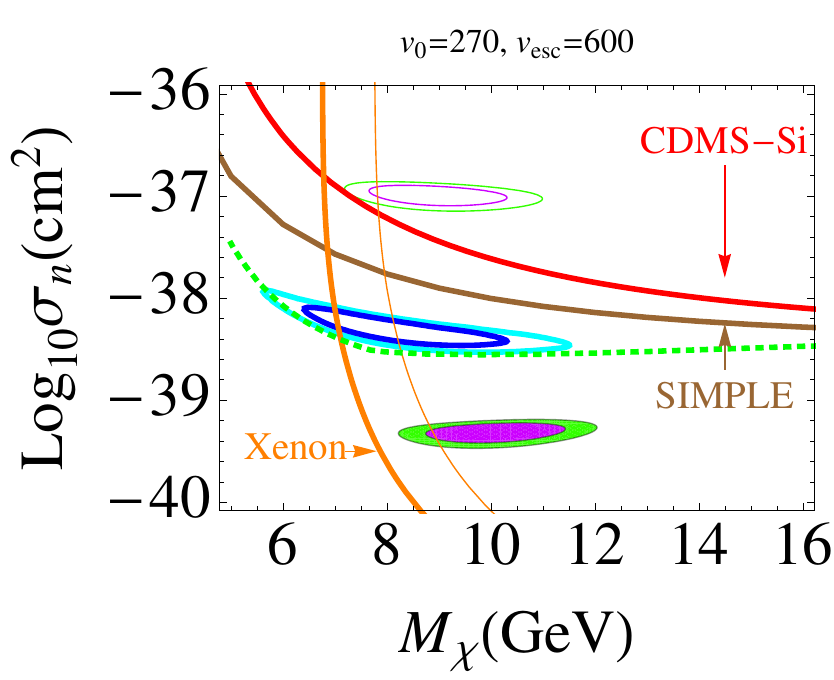}
\end{center}
\caption{Spin independent scattering with $f_n = -f_p$.  Contours and lines are labeled as in figure \ref{fig:nobgcontours}, with the addition of the XENON10 limit from varying $\mathcal{L}_{eff}$ at 1$\sigma$ (orange thin).}
\label{fig:nobgcontoursfneqmfp}
\end{figure}

In Fig.~\ref{fig:nobgcontoursfneqmfp}, we show the signal regions and existing constraints for the $f_{p} =-f_n$ case for four different velocity distributions.  As expected, the constraint from Si falls away.  The constraint from XENON10 is still relevant.  It would excise all but the lowest masses using the central value of the $\mathcal{L}_{eff}$ measurements of \cite{Manzur:2009hp}.  Similar to what was done in \cite{Fitzpatrick:2010em}, we also show a curve that would be derived by varying the $\mathcal{L}_{eff}$ measurements within one-sigma below the central  measured values.  This opens more parameter space near 10 GeV.  Especially for the largest values of $v_{0}$, it appears that a lower value of  $\mathcal{L}_{eff}$ might potentially allow coincidence between DAMA and CoGeNT in a region not previously excluded.  Examining the lower-left plot, shifting DAMA upwards (as would be appropriate, should the canonical estimates of the channeling fraction be too large), the DAMA and CoGeNT regions can overlap.

\begin{figure}[h]
\begin{center}
\includegraphics[width=0.45\textwidth]{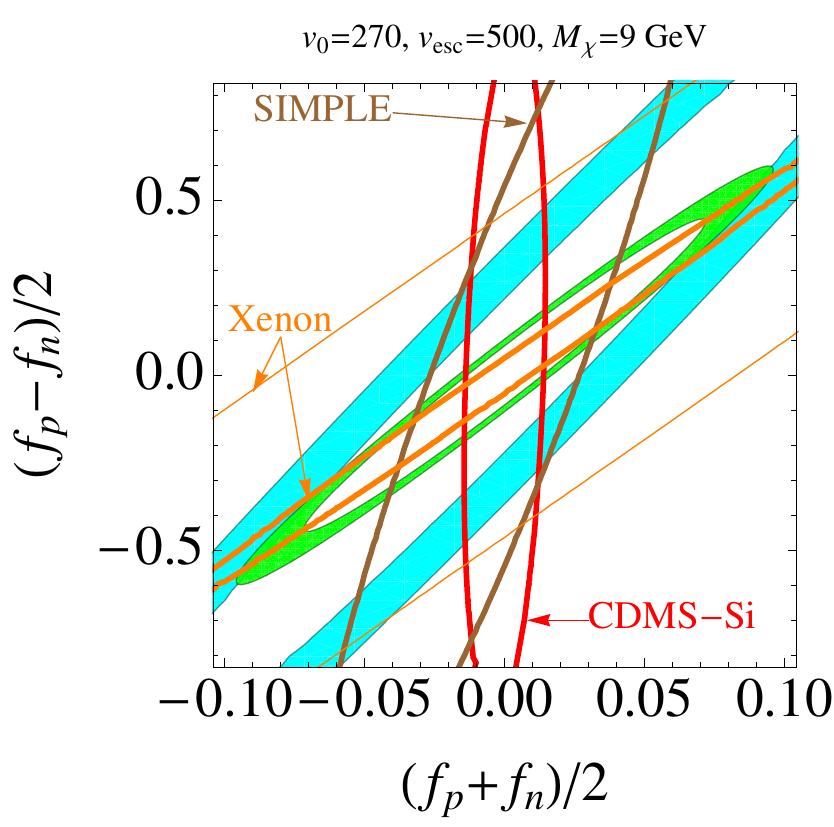}
\includegraphics[width=0.45\textwidth]{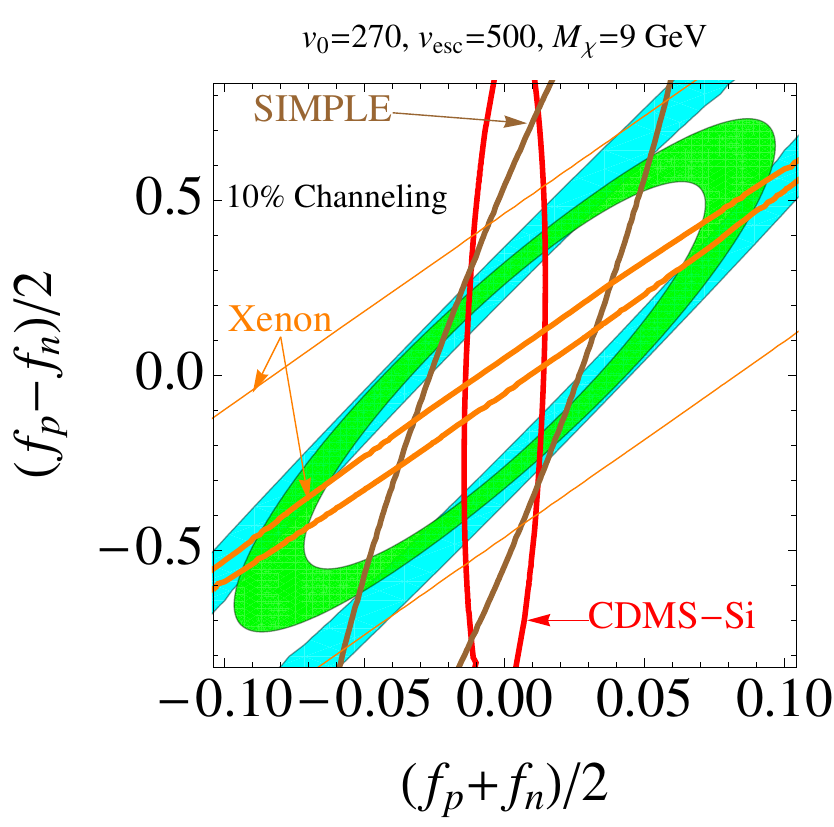}\\
\includegraphics[width=0.45\textwidth]{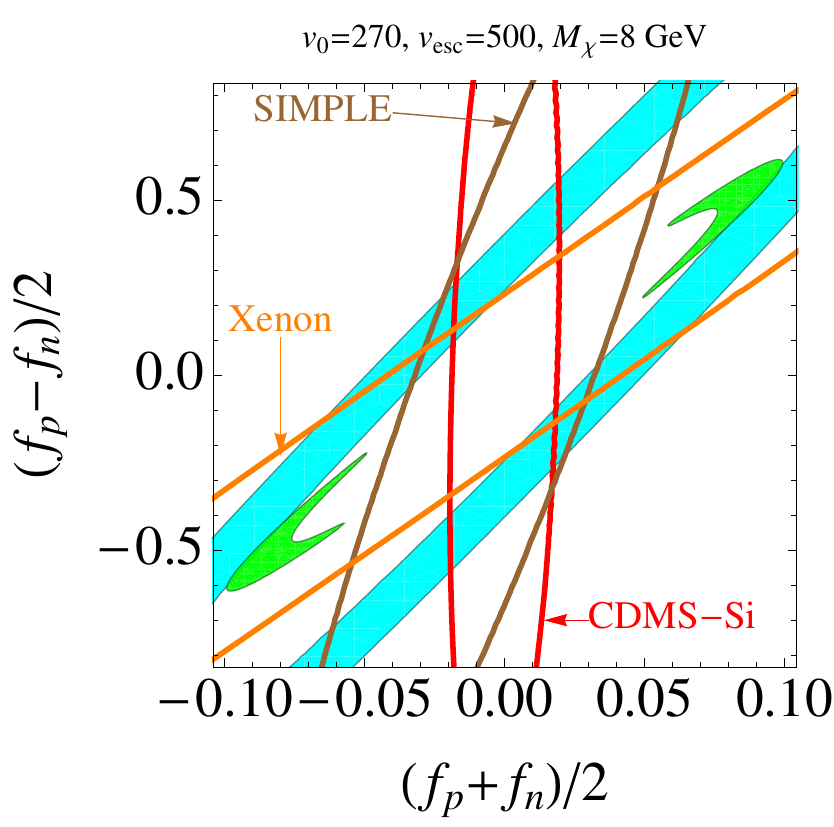}
\end{center}
\caption{Generalized proton and neutron coupling plots for SI interactions.  The limits are CDMS-Si (red), SIMPLE (brown), and XENON10 (orange thick and thin lines for $\mathcal{L}_{eff}$ central value and 1$\sigma$ variation).  Filled contours are 99\% CL goodness of fit for CoGeNT (cyan) and DAMA (green).  Top plots are for 9 GeV, the left plot with  normal channeling and the right plot with 10\% of channeling.  Bottom:  Normal channeling with 8 GeV mass.  The only visible XENON10 limit is the central value one.  Spectral constraints from DAMA impose that the couplings are only consistent with iodine suppression ($f_n \approx -.7 f_p$).}
\label{fig:fsvsfaplot}
\end{figure}

In Fig.~\ref{fig:fsvsfaplot}, we show the region of allowed couplings in the $(f_{p}-f_{n})/2$ vs. $(f_{p}+f_{n})/2$ plane, for the optimistic case of $v_0 =270$, $v_{esc}=500$.  For comparison with previous plots, the cross section per nucleon $\sigma_{(p,n)} \approx (5.9 f_{(p,n)}^2)\times 10^{-38} {\rm cm}^2$. The constraints from CDMS Si, SIMPLE, and Xenon show up as ellipses.  The constraints force the couplings to lie interior to each ellipse.  Signal regions show up as the filled-in elliptical generalizations of annuli, where one is constrained to live within the filled regions.  These regions represent a 99\% CL goodness of fit, which we use for the following reason.  The type of models which would realize these different relative strengths in couplings can come from different physics scenarios. It does not seem suitable to perform a $\Delta \chi^2$ analysis around the best fit point in these plots, as this would be comparing different theories. In contrast, the goodness of fit regions give a more model-independent determination of the allowed parameter space.  Additionally, we chose to only plot the 99\% contour and not both the 90\% and 99\% regions (as done for previous signal regions) for clarity, as these two regions are similar. Again, we do not include an exponential background.  Given that there are 24 (7) degrees of freedom for the CoGeNT (DAMA) $\chi^2$, these regions correspond to $\chi^2< 42.97\; (18.48)$.

Looking at the top left plot, we see that for a 9 GeV WIMP, the DAMA and CoGeNT preferred regions has aside from XENON10's central limit, no constraints between them when $f_n=-f_p$.  Reducing the channeling for both sodium and iodine to 10\% of the prediction of \cite{Bernabei:2007hw}, in the top right plot, we see that we can get consistency between the CoGeNT and DAMA regions.  Whether such consistency can be obtained depends strongly on the Dark Matter mass. At light masses (below 10 GeV), the spectral shape observed at DAMA prefers a coupling to sodium over iodine. The impact of this effect can be seen in the bottom plot of  Fig.~\ref{fig:fsvsfaplot}: for a 8 GeV WIMP, the contours that fit DAMA are in a region where the coupling to iodine is effectively cancelled, $f_n \approx -.7 f_p$ (equivalently $(f_p-f_n) \approx 6.7 (f_p+f_n)$).    At 8 GeV, the XENON10 constraint with the central value of $\mathcal{L}_{eff}$ is still a strong constraint.  The DAMA spectrum prevents going to even lower masses where the XENON10 constraint would no longer be relevant.

Modifications of how we analyze DAMA and CoGeNT can open up this region, e.g., increasing the uncertainty for the first bin of DAMA or adding in some of the exponential background into CoGeNT.  Although we have not performed a systematic study of varying the amount of channeling separately for sodium and iodine, we were unable to find values which worked for such a light WIMP.   Finally, if one goes to lower $v_0$ or higher $v_{esc}$, the constraints from XENON10 become more severe and the mass range where consistency can be obtained between DAMA and CoGeNT gets smaller.
\subsection{Spin Dependent Elastic Scattering}
\begin{figure}[h]
\begin{center}
\includegraphics[width=0.45\textwidth]{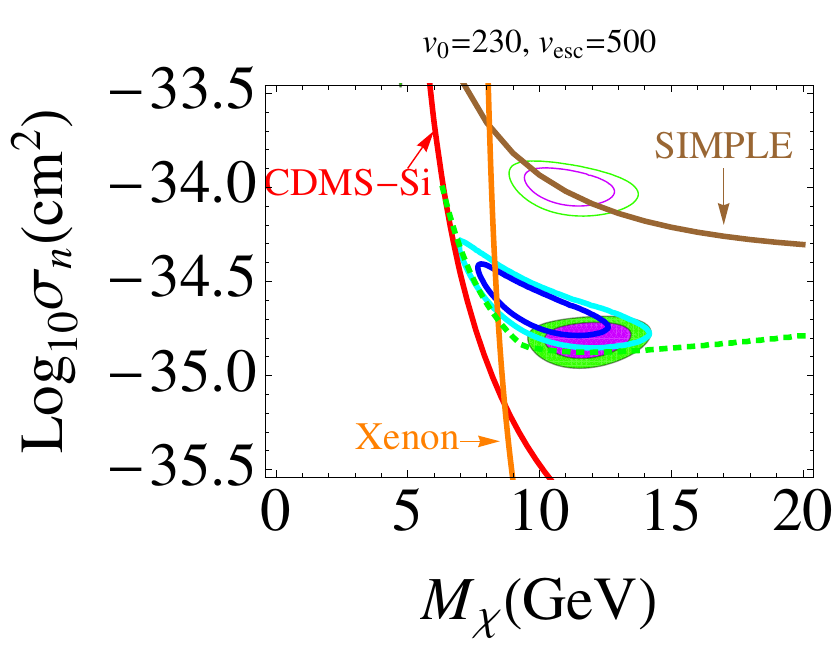}\quad
\includegraphics[width=0.45\textwidth]{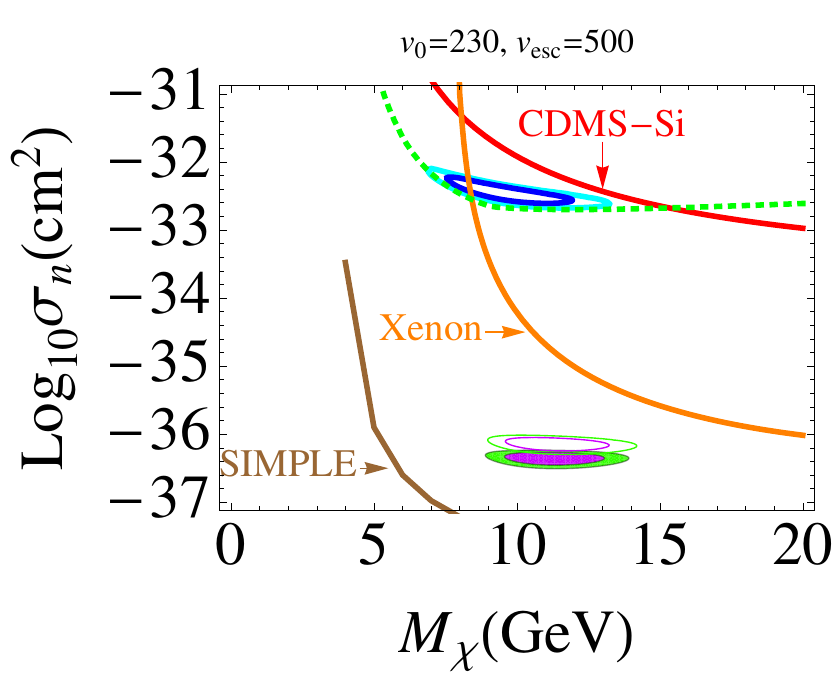}
\end{center}
\caption{Left plot is spin dependent neutron scattering, right plot is spin dependent proton scattering.  Labeling is as in figure \ref{fig:nobgcontours}.}
\label{fig:SDplot}
\end{figure}
Spin dependent interactions are a logical next step in terms of potential CoGeNT explanations, such an analysis appeared recently in   
\cite{Kopp:2009qt}.  Consistent with their results, we find that in the case where the Dark Matter couples exclusively to neutron spin, the DAMA and CoGeNT regions overlap, see Fig.~\ref{fig:SDplot}.  However, these regions are excluded by CDMS-Si.  
The constraint from XENON10 is also important.    In Fig.~\ref{fig:SDplot}, we also show the regions when the Dark Matter couples to proton spin.  In this case, there are several experiments that exclude this interpretation of the CoGeNT data.  For example, the SIMPLE experiment (shown) excludes the CoGeNT cross section by over five orders of magnitude.

\begin{figure}[h]
\begin{center}
\includegraphics[width=0.45\textwidth]{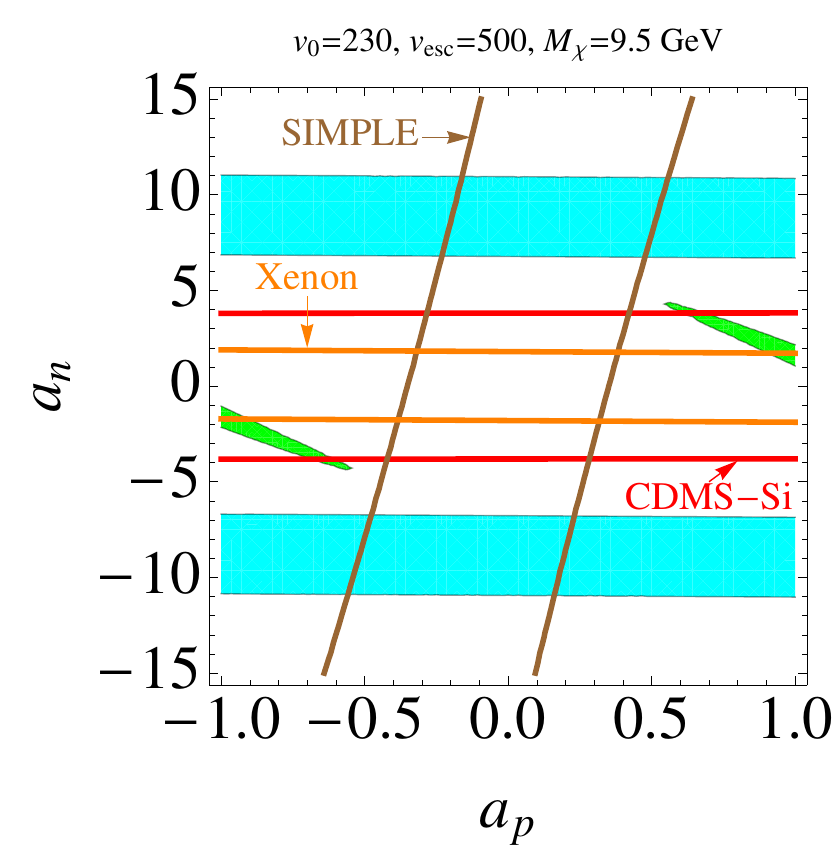}\quad
\includegraphics[width=0.45\textwidth]{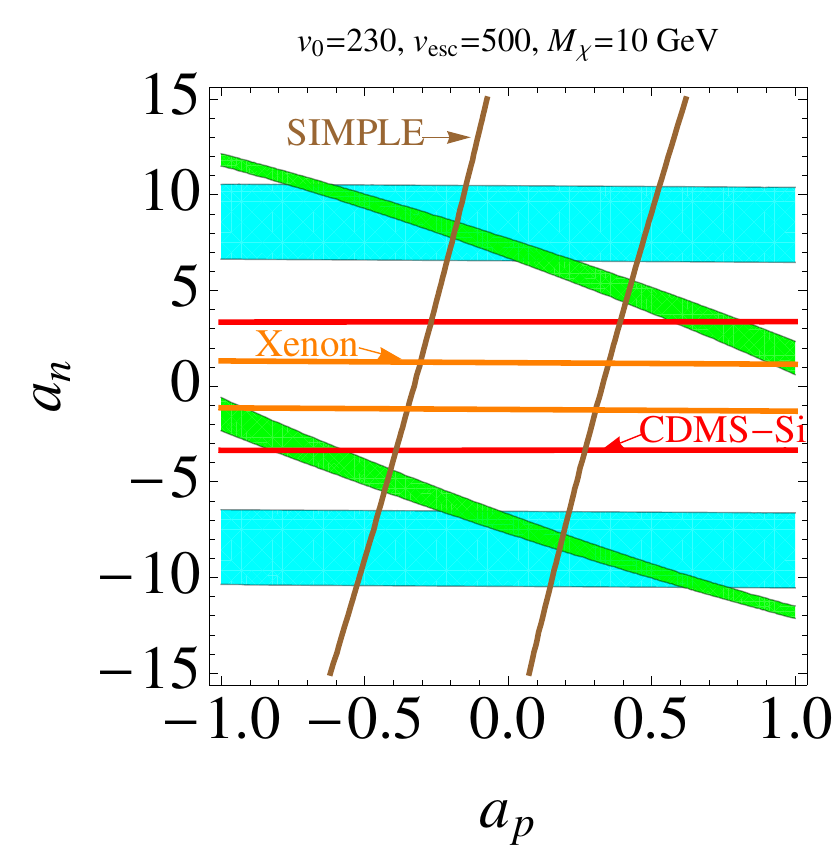}
\end{center}
\caption{Generalized spin dependent proton and neutron couplings for 9.5 and 10 GeV WIMPs.  Only the XENON10 central limit is displayed on both plots.  Labeling is as in figure \ref{fig:nobgcontoursfneqmfp}. }
\label{fig:apvsanplot}
\end{figure}

Unfortunately, unlike the case of spin independent interactions, generalized couplings for protons and neutrons do not open additional parameter space. Again, to compare with previous plots, the cross section per nucleon $\sigma_{(p,n)} \approx (3.6 a_{(p,n)}^2)\times 10^{-37} {\rm cm}^2$. In Fig.~\ref{fig:apvsanplot}, WIMPs of 9.5 and 10 GeV have significant constraints from SIMPLE and CDMS-Si which rule out all of the CoGeNT and DAMA preferred regions.  As was seen in the generalized spin independent interactions,  the DAMA fit at low WIMP masses is again preferring sodium over iodine scattering, choosing the relative proton and neutron couplings to reduce the iodine contribution.    

\subsection{Momentum Dependent Dark Matter}
As recently emphasized in \cite{MDDM}, there are dimension six operators that can contribute to direct detection beyond the usual ones relevant for SI and SD scattering.  Specifically,
\begin{eqnarray}
{\mathcal O}_{1}&=&(\bar{\chi} \gamma _{5} \chi) (\bar q   q), \label{eqn:operators1} \\
{\mathcal O}_{2}&=&(\bar{\chi}  \chi) (\bar q  \gamma_5 q), \\
{\mathcal O}_{3}&=&(\bar{\chi} \gamma_{5} \chi) ( \bar q  \gamma_{5} q), \\
{\mathcal O}_{4}&=&(\bar{\chi} \gamma_{\mu} \gamma_{5} \chi) (\bar{q} \gamma^{\mu} q).
\label{eqn:operators}
\end{eqnarray}
These operators suffer from suppression by either the velocity of the DM ($v_{\rm DM} \sim 10^{-3}c$) or $q^2$, but this can be compensated if the mediator (i.e. the particle integrated out to obtain the operator) is light.  The first three operators are identical to either SI or SD operators, modified only by powers of $q^{2}= 2M_{N} E_{R}$ when compared to the usual operators.  This dependence  on the nucleon mass can impact the relative power of various experiments.  Similar non-minimal scattering was considered in \cite{FFDM}, where dark gauge bosons coupled to magnetic moment operators of the DM.  Kinetic mixing of these dark gauge bosons with the photon, leads to momentum suppressed SI interactions that couple to the charge of the nucleus.   Here, we take a phenomenological approach, allowing generalized couplings and momentum dependence, allowing powers up to $q^4$.  

Momentum dependent interactions give an increased rate at CoGeNT when compared with CDMS-Si, alleviating some tension there.  On the other hand, such interactions would not generically help find a common region for DAMA and CoGeNT that is consistent with other experiments, as they enhance the rate of XENON10 vs  CoGeNT.  However,  the momentum dependence improves the fits of lower mass DM to the spectra observed at DAMA and CoGeNT. At very low masses, it is hard to scatter at XENON10 due to the escape velocity cutoff, so in certain special cases a more generalized $q^{2}$ dependence allows consistency between all experiments.  

\begin{figure}[h]
\begin{center}
\includegraphics[width=0.45\textwidth]{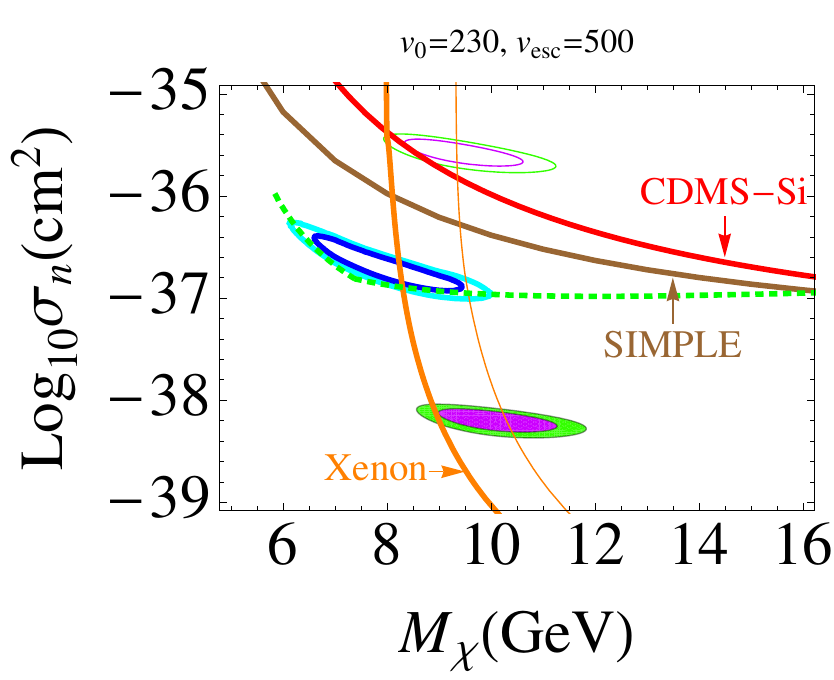}
\includegraphics[width=0.45\textwidth]{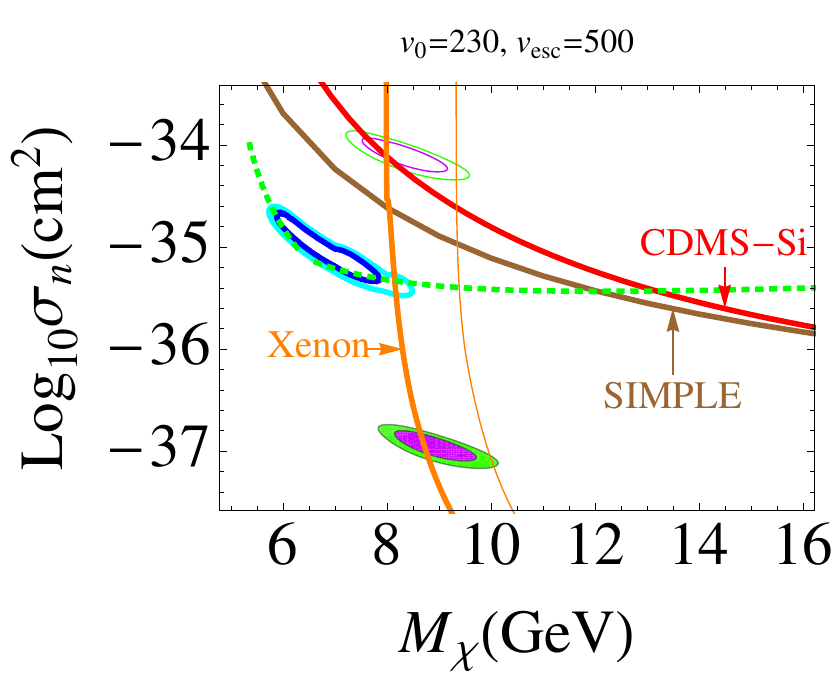}\\
\end{center}
\caption{Spin independent scattering with $f_n = -f_p$ with $q^2$ and $q^4$ momentum dependence on the left and right.  Labeling is as in figure \ref{fig:nobgcontours}.}
\label{fig:nobgcontoursfneqmfpmddm}
\end{figure}

\begin{figure}[h]
\begin{center}
\includegraphics[width=0.45\textwidth]{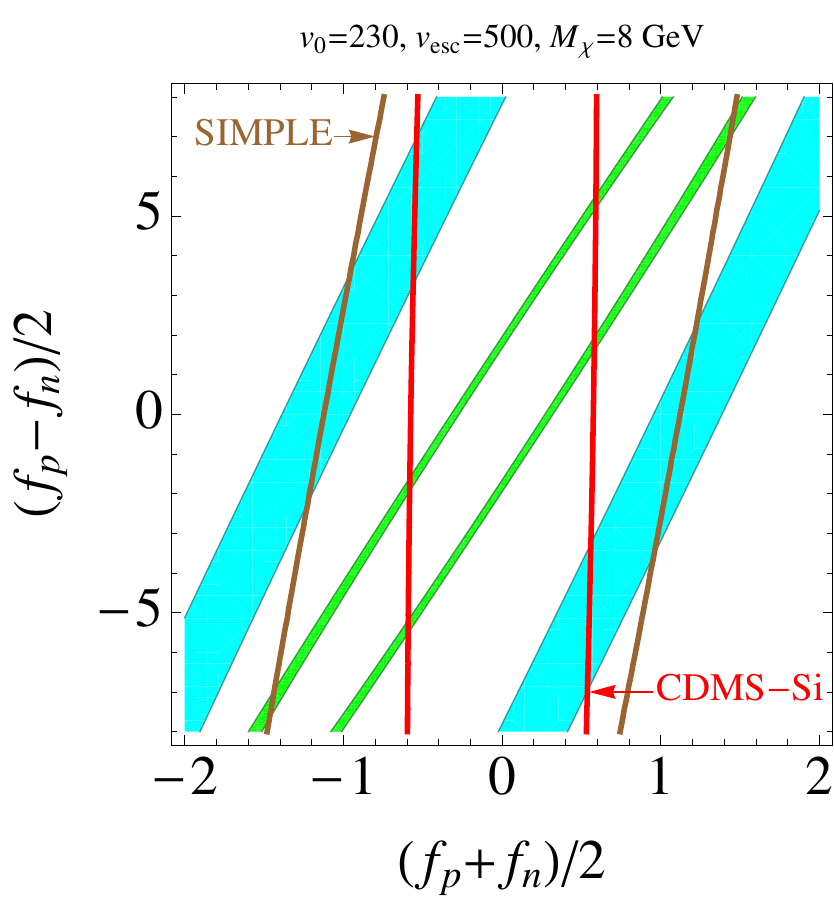}
\includegraphics[width=0.45\textwidth]{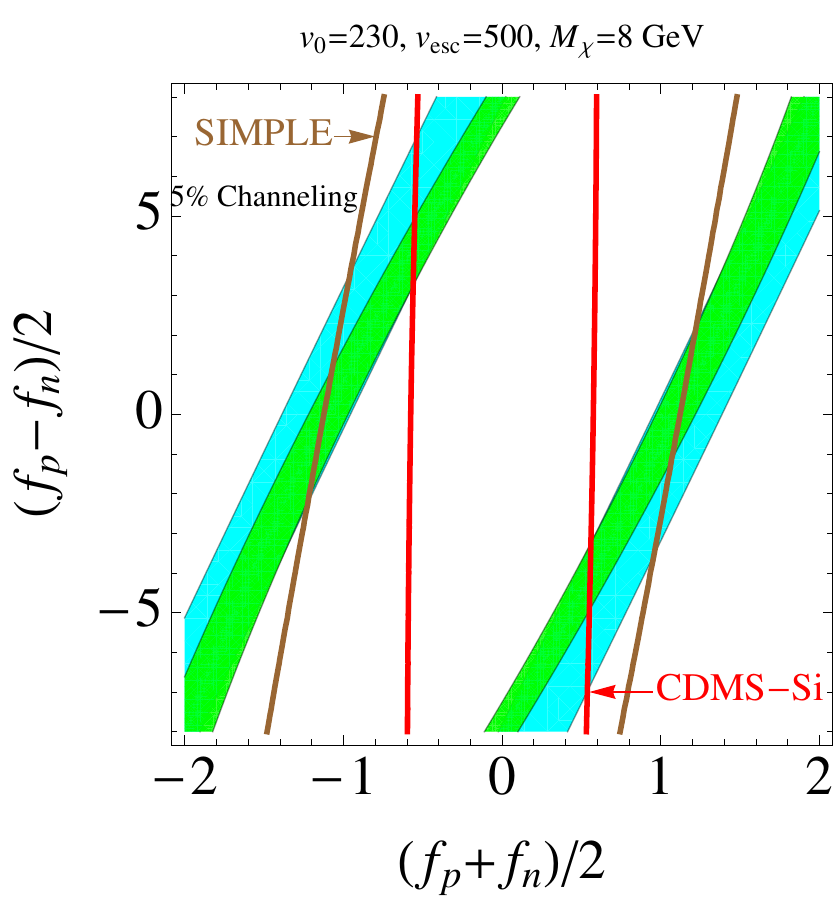}\\
\end{center}
\caption{
Generalized spin independent couplings with a $q^4$ momentum dependence for 8 GeV DM.  Left plot has normal channeling, right plot has 5\% of channeling.  Labeling is as in figure \ref{fig:nobgcontoursfneqmfp}.}
\label{fig:fsvsfaplotmddm}
\end{figure}

For example, in Fig.~\ref{fig:nobgcontoursfneqmfpmddm}, we have plotted the $f_n=-f_p$ SI scenario with a momentum dependence of $q^2/(100\MeV)^2$ and  $q^4/(100\MeV)^4$.  Compared to the upper left plot of Fig.~\ref{fig:nobgcontoursfneqmfp}, we see that both the CoGeNT and DAMA regions move to lower masses, with CoGeNT having the larger shift.  At CoGeNT, the momentum suppression at low energy gives a softer exponential, thus mimicking heavier masses.  At DAMA, the additional powers of momentum suppress the low energy scattering for iodine and sodium, allowing lighter masses to be consistent with the first two bins of DAMA's modulation spectra.  Without this additional suppression, light DM overpredicts these bins.  Generalizing couplings to protons and neutrons, we see in Fig.~\ref{fig:fsvsfaplotmddm} that for 8 GeV DM, the XENON10 central limit is no longer constraining and if channeling is only 5\% of the benchmark suggested  by the DAMA collaboration, the DAMA and CoGeNT regions agree.

Adding in momentum dependence for SD couplings improves the agreement of the signal regions with the null results, especially for CoGeNT, but does not allow regions consistent for both DAMA and CoGeNT.  For SD couplings to neutrons, the DAMA region moves to lighter masses, but is still ruled out by CDMS-Si.  However, a small sliver of CoGeNT opens up at low mass where it can avoid the CDMS-Si limit, due to the $q^2$ suppressing silicon scattering more than germanium.  For SD couplings to protons, germanium does not couple strongly to protons, so the CoGeNT region is much higher than the DAMA region, where the SIMPLE constraints easily rule it out.  For this case, DAMA is also ruled out by SIMPLE, since the DAMA region does not move much given the substantial sodium coupling to protons.  Also, we find generalized SD couplings do not help significantly since experiments are roughly ``orthogonal'' in sensitivities to proton and neutron couplings, as seen in Fig.~\ref{fig:apvsanplot}.  It does allow DAMA regions that are only ruled out by the XENON10 central limit, but it doesn't create regions which are consistent with DAMA and CoGeNT at the same time.  Therefore, even though momentum dependent SD couplings are in less sharp conflict with the data, unlike the SI case, it does not open up significant regions of parameter space.          

\begin{figure}[h]
\begin{center}
\includegraphics[width=0.6\textwidth]{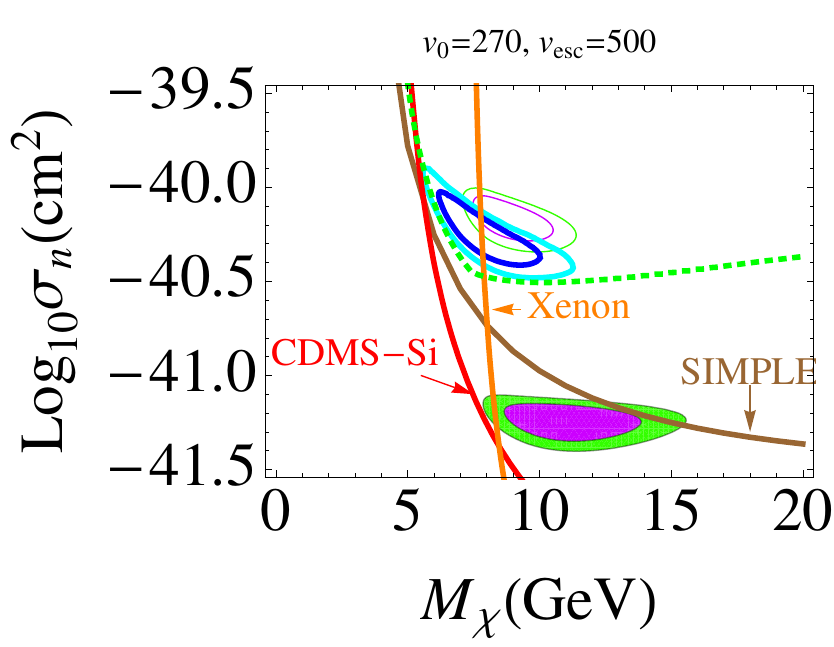}
\end{center}
\caption{Anapole scattering with scattering to the charge of the nucleus, with labeling is as in figure \ref{fig:nobgcontours}.}
\label{fig:anapole}
\end{figure}

We also comment on ``anapole'' scattering given by ${\mathcal O}_{4}$ in Eq.~(\ref{eqn:operators}).  Not only does it have a $q^{2}$ piece, but it also has a piece that goes like two additional powers of velocity.  
The kinematics give a different recoil formula, which to leading order in $q^2$ and $v^2$ we normalize as
\begin{eqnarray}
\frac{dR_{anapole}}{dE_R}= \frac{N_T m_N \rho_\chi}{2m_\chi \mu^2} \left(\frac{c}{v_0}\right)^2\sigma(q^2) \int^\infty_{v_{min}} \left[4 \left(\frac{v}{c}\right)^2-q^2\left(\frac{1}{m_\chi^2}+\frac{2}{m_\chi m_N}-\frac{1}{m_N^2}\right)\right]\frac{f(v)}{v}dv. \nonumber \\[-.1cm]
&& 
\end{eqnarray}
This operator suffers from suppression from the factor in brackets inside the integral, but this can be compensated if the mediator is light.  In fact, this operator naturally arises if there is a light gauge boson (weakly coupled to the standard model) that couples to a Majorana DM fermion.   If there is kinetic mixing with the photon, this couples dominantly to the charge of the nucleus.  The allowed region for this operator is shown in Fig. \ref{fig:anapole}.  Again, there is significant tension with SIMPLE and CDMS-Si, with SIMPLE improving relative to XENON10, due to the two factors in the integrand canceling more for heavy targets like xenon.    With respect to the corresponding SI model (see Fig.~\ref{fig:nobgcontours}), the anapole allowed region is in milder conflict with extant experiments.  Amusingly, the CoGeNT region overlaps with the unchanneled DAMA region, but the region of overlap appears naively inconsistent with both CDMS-Si and SIMPLE.

\subsection{Inelastic Dark Matter}
\begin{figure}[h]
 \begin{center}
\includegraphics[scale=0.92]{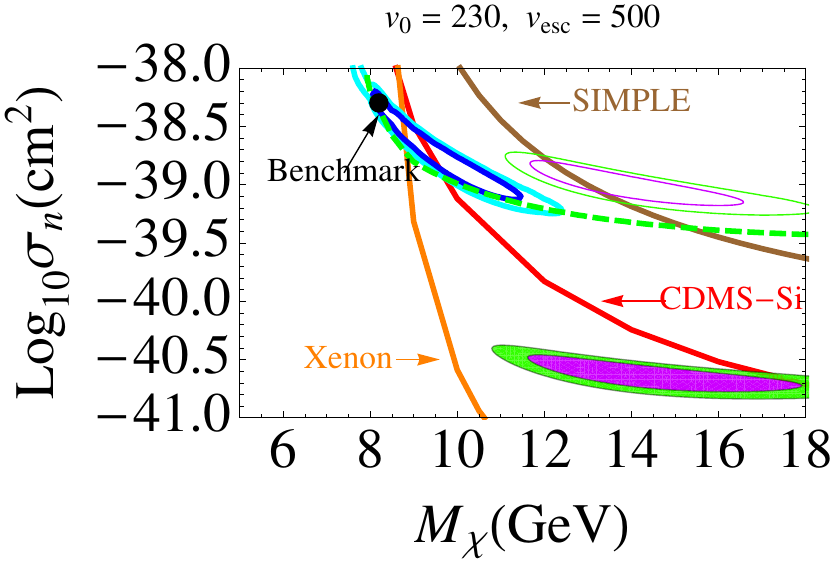}
\includegraphics[width=0.45\textwidth]{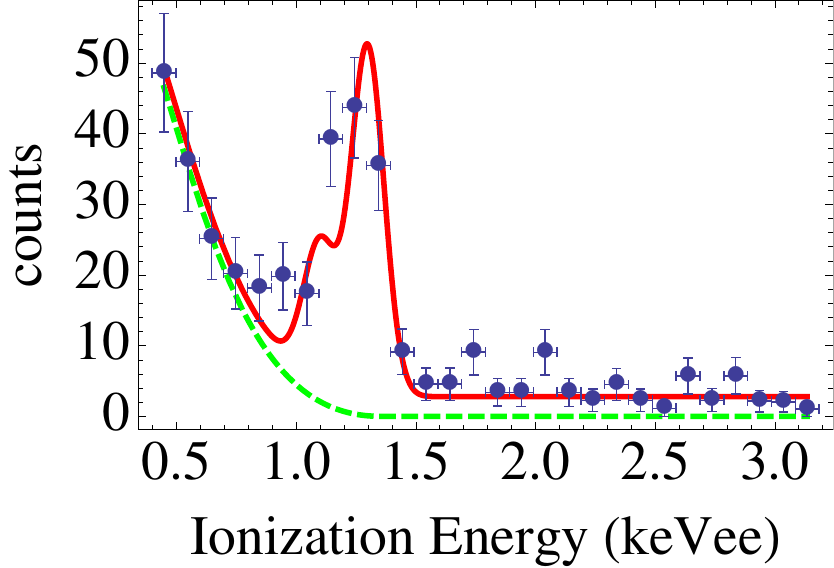}
\end{center}
\caption{On the left pane we plot the parameter space for inelastic DM with splitting of $\delta = 15$ keV.  Labeling is as in figure \ref{fig:nobgcontours}. The spectrum for a benchmark point (indicated in the figure) is shown in the right pane ($\mX = 8.2\GeV$, $\sX = 10^{-38.3}\cm^2$). }
\label{fig:inel-expbgcontours}
\end{figure}

One final scenario in which the relative signals are modified is inelastic DM~\cite{iDM}. Here the DM sector consists of at least two states, $\chi$ and $\chi^{\ast}$, with masses split by $\delta \equiv m_{\chi^{\ast}}- m_{\chi}$. The minimum velocity required to recoil with deposited energy, $\ER$, is given by
\begin{equation}
v_{\rm min} = \frac{1}{\sqrt{2\mN\ER}}\left(\frac{\mN\ER}{\mu} + \delta \right).
\end{equation}
Therefore, experiments with heavier targets have greater sensitivity to inelastic transitions (see \cite{iDM} for more details). In particular, the signal event rate in CDMS-Si, which is the most stringent constraint on the elastic spin-independent case, is suppressed as compared to CoGeNT and opens up regions of parameter space not excluded by other experiments. This is shown in Fig.~\ref{fig:inel-expbgcontours} where we plot the $90\%$ CL contours in the $\mX$-$\sX$ plane for inelastic scattering with $\delta = 15\keV$ as well as constraints from CDMS-Si, XENON10, and SIMPLE. As can be expected from kinematics, the strongest constraints  come from XENON10. The small part of parameter space which is still viable at low WIMP mass ($\mX \lesssim 9\GeV$) is not consistent with the DAMA results either with or without channeling. 
On the right pane of Fig.~\ref{fig:inel-expbgcontours} we plot the spectrum for a particular point in the allowed region of parameter space.  Notice that the inelastic splitting is small enough that there is little suppression of the low energy events at CoGeNT.  This is contrary to the expectations of 100 keV splittings \cite{iDM}.
  
SD inelastic interactions can change the relative signals at different experiments, as recently pointed out in \cite{Kopp:2009qt}. We consider this possibility for CoGeNT and/or DAMA. For SD neutron scattering,  the CDMS-Si and XENON10 constraints can be weakened by increasing the inelastic mass splitting so as to allow consistent regions at low mass for CoGeNT and DAMA separately. 
However, as the mass splitting increases to about $25\keV$, CoGeNT is not fit well by DM as the low energy bins are suppressed; this ensures that  the CDMS-Si limit always separates the CoGeNT and DAMA regions.  Thus, the exponential background at CoGeNT will have to be introduced, to get consistency between the two regions.  For SD proton scattering, again it is possible to get consistent DAMA and CoGeNT regions separately, however, it is hard to bridge the gulf between the required cross sections for CoGeNT and DAMA, since the CoGeNT cross section is much larger than the no channeling region for DAMA.  Thus, it is even more unlikely to have a consistent region where both experiments are explained for SD proton couplings.  Allowing generalized spin couplings, we find that it is possible to get the CoGeNT and DAMA regions closer, but due to CDMS-Si and XENON10 limits, the CoGeNT region again must be lowered to get consistent regions. 
  
In summary,  for both SI and SD inelastic scattering of DM, it is possible to reduce limits for CoGeNT at low mass, but difficult to get consistency between CoGeNT and DAMA without introducing a substantial exponential background into CoGeNT.    

\subsection{Models \label{sec:models}}
In all cases considered, the required cross sections are quite large -  several orders of magnitude larger than would be expected by Higgs boson exchange - and the masses well below a wide range of existing collider experiments. It is worthwhile to comment on how such a light particle might both freeze out with the appropriate cross section and have large couplings to standard model matter.

If DM $\chi$ freezes out by annihilating into a new, light ($\lsim\GeV$) boson $\phi$, the ``WIMP miracle'' can be preserved, if the $\phi$ maintains equilibrium with the standard model \cite{Finkbeiner:2007kk,Pospelov:2007mp}. Such a freezeout process has been employed to construct models of MeV DM \cite{Pospelov:2007mp}. More recently, such scenarios have been employed to create light, but still weak-scale models \cite{Zurek:2008qg,Morrissey:2009ur} to explain DAMA through a $\sim$ 10 GeV WIMP. This basic setup allows us to realize most of the models we have considered.

We will begin by considering a supersymmetrized $U(1)_d$. The connection between the dark sector and the standard model arises through a $U(1)_{EM}-U(1)_d$ kinetic mixing, which is easily supersymmetrized, i.e.
\be
{\cal L} \supset \epsilon F^{Y}_{\mu \nu} F^{\mu \nu}_d \rightarrow \epsilon  \int d^2 \theta W^{Y}_\alpha W^d_\alpha.
\ee
This latter term generates a D-term vev when electroweak symmetry is broken \cite{Cheung:2009qd}. Although the interaction strength is proportional to $\epsilon \ll 1$, so, too, is the mass squared. Under such circumstances, a cancellation occurs. If we assume that the dark sector contains a stable field vectorlike under the $U(1)_d$, it will have  a natural range for its scattering cross section $10^{-41} \rm{cm^2} \lsim \sigma_0 \lsim 10^{-37} \rm{cm^2}$ \cite{Cheung:2009qd}.  Thus one can achieve a large scattering cross section, without concern of the production of the WIMP in collider experiments\footnote{For a review of the limits on such particles, see the discussion in \cite{Essig:2009nc,Bjorken:2009mm}.}.

If the Dirac fermion that constitutes DM is perturbed by a small Majorana mass after $U(1)_d$ breaking (or small scalar-pseudoscalar splitting, in the case of a complex scalar), the interaction will be off-diagonal. I.e., the Dirac fermion can be decomposed as two approximately degenerate Majorana fermions, between whom the $U(1)$ mediates an inelastic transition. Diagonalizing the kinetic terms can naturally generate a Majorana mass term $\epsilon^2 M_{1}$, although it is also possible that in gauge mediated models the same rotation will diagonalize both terms. Other higher-order corrections can generate the same effect. Such a setup then realizes a light version of the inelastic DM models considered in \cite{ArkaniHamed:2008qn,Baumgart:2009tn,Cui:2009xq,Katz:2009qq,Morrissey:2009ur}. 

If the Majorana mass is too large to allow inelastic scattering, a large anapole operator is present instead. While the natural scale of this interaction is typically $v^2/c^2$ smaller than a spin-independent vector interaction, with some tuning of masses it is possible to achieve a value relevant for scattering experiments.

It is more challenging to generate the $f_p = -f_n$ scenario. It is possible to induce different couplings to protons and neutrons in multi-Higgs boson models, but it is difficult to understand how $f_p = -f_n$ would arise. This is especially true when one accounts for the common strange (and heavy quark) content in the two nucleons. These heavier quarks typically dominate the direct detection cross section. It is possible that a coupling to a light scalar (coupling more universally to protons and neutrons) could interfere with an interaction mediated by a dark photon, but this would require a very severe coincidence. Alternatively, if a sizeable interaction with the $\rho$-meson is present, DM could couple dominantly to isospin, which would realize the $f_p=-f_n$ scenario. We defer a detailed examination of these possibilities to future work. 

\section{Conclusions \label{sec:conclusions}}
The recent CoGeNT results are intriguing, both because of the lack of an obvious background, and the possible connection to the DAMA result.
In this paper, we have taken the CoGeNT results and asked the question, in what cases can DM interactions consistently explain a substantial number of their lower energy events?  The limits of XENON10, SIMPLE, and especially CDMS-Si give significant constraints. Taken together, they make it difficult to answer this question in the affirmative without considering somewhat unconventional scenarios.  To summarize our results:      

\begin{itemize}
\item An interpretation of the low energy excess events as coming entirely from elastic scattering, either spin-independent or spin-dependent, of WIMPs against nucleons is possible and results in a good fit to the data. However, the implied parameter space is strongly disfavored by other searches.

\item Postulating an additional background component at low energies helps to avoid existing constraints by allowing a reasonable fit to the data with lower WIMP mass and/or lower WIMP-nucleon cross-section. However, the WIMP signal in these regions of parameter space cannot be said to explain the data as it contributes less (and generally substantially less) than 50\% to at least one of the first 5 bins. At  best one can say that a WIMP signal is not inconsistent with the data if we allow additional background component at low energies.  

\item  Modifications beyond standard scattering can improve the situation. 
\begin{itemize}
\item Light inelastic DM with SI or SD interactions can explain the low energy events in the CoGeNT data, but have difficulty explaining both DAMA and CoGeNT with DM alone.
\item Generalized SI couplings to protons and neutrons are the most promising, in particular when $f_p\approx -f_n$. This is especially true when considered in conjunction with momentum dependence. This allows situations where DM is simultaneously responsible for the CoGeNT and DAMA signals.
\end{itemize}    
\end{itemize}

We leave it for future work to construct models which explain the necessary couplings of these nonstandard scattering mechanisms.  In the near future, further results from CDMS, from either additional silicon data or a low threshold germanium analysis, could impact whether these scenarios are still viable. More optimistically,  a confirmation of these light mass WIMPs could be coming soon from COUPP, XENON100, XMASS, and LUX, depending on their low energy threshold.  

\vskip 0.15in
\textbf{Note Added:} We would like to thank the authors of Refs.~\cite{slac-paper,Graham:2010ca} for bringing to our attention their related work. Our analysis of the elastic spin-independent constraints are in agreement with what was found in \cite{Graham:2010ca} except in the limits coming from the CDMS-Si runs. Ref.~\cite{Graham:2010ca} used in addition the results of CDMS-SUF~\cite{Akerib:2003px} which we have omitted for two reasons. First, CDMS-SUF was a fairly different experiment with larger backgrounds (not shielded like in CDMS-Soudan). Second, it had a much smaller exposure and most of its power is derived from the low threshold ($5\keV$) where the efficiency is uncertain. To remain conservative in our limits we opted to include only the 2-tower data \cite{Akerib:2005kh} and the 5-tower data shown in \cite{Filippini}.

\acknowledgments

We would like to especially thank Juan Collar for helping us understand the CoGeNT data and analysis.  We would like to thank Kyle Cranmer for insightful conversations.  AP would also like to thank T. Cohen, D. Phalen and K. Zurek  for related discussions. The work of SC is supported under DOE Grant \#DE-FG02-91ER40674.  The work of A.P. was supported in part by NSF Career Grant NSF-PHY-0743315 and by DOE Grant \#DE-FG02-95ER40899. NW is supported by Department of Energy OJI grant \#~DE-FG02-06ER41417.  I. Y. is supported by the James Arthur fellowship. 

\bibliographystyle{JHEP}
\bibliography{cogent}
\end{document}